\documentclass[10pt,letterpaper]{article}

\usepackage{graphicx}
\usepackage{url}
\usepackage{color}
\usepackage{array}
\usepackage{comment}

\usepackage{lscape}
\usepackage{rotating}

 \usepackage[pdftex,colorlinks=true, linkcolor=blue,citecolor=blue,
       urlcolor=blue]{hyperref}

\title{Many-Task Computing and Blue Waters\footnote{Please cite as: D. S. Katz, T. G. Armstrong, Z. Zhang, M. Wilde, and J. M. Wozniak,  Many Task Computing and Blue Waters. Technical Report CI-TR-13-0911. Computation Institute, University of Chicago \& Argonne National Laboratory. \url{http://www.ci.uchicago.edu/research/papers/CI-TR-13-0911}}}

\author{Daniel S. Katz,$^{1}$ Timothy G. Armstrong,$^{2}$ Zhao Zhang,$^{2}$\\
Michael Wilde,$^{1,3}$ Justin M. Wozniak$^{1,3}$\\
  \small $^1$Computation Institute,
  University of Chicago and Argonne National Laboratory\\[-0.2em]
  \small $^2$Department of Computer Science,
  University of Chicago\\[-0.2em]
  \small $^3$Mathematics and Computer Science Division,
  Argonne National Laboratory\\[-0.2em]
}

\date{February 13, 2012}

\begin{document}

\maketitle

\begin{abstract}
This report discusses many-task computing (MTC), both generically and
in the context of the proposed Blue Waters systems.  Blue Waters is planned to be the largest
supercomputer funded by NSF when
it begins production use in 2011--2012 at NCSA.
The aim of this report is to inform the Blue Waters project about MTC,
including understanding aspects of MTC applications that can be used to
characterize the domain and understanding the implications of these
aspects to middleware and policies on Blue Waters.

Many MTC applications do not neatly fit the stereotypes of
high-performance computing (HPC) or high-throughput computing (HTC)
applications.  Like HTC applications, by definition MTC applications
are structured as graphs of discrete tasks, with explicit input and output
dependencies forming the graph edges.  However, MTC applications have
significant features that distinguish them from typical HTC applications.
In particular, different engineering constraints for hardware and software must be
met in order to support these applications.

HTC applications have traditionally run on platforms such as grids and clusters,
through either workflow systems or parallel programming systems.
MTC applications, in contrast, will
often demand a short time to solution, may be communication intensive or
data intensive, and may comprise very short tasks.
Therefore, hardware and software for MTC must be engineered
to support the additional communication and I/O and must minimize
task dispatch overheads.

The hardware of large-scale HPC systems such as Blue Waters, with its high degree of
parallelism and support for intensive communication, is well suited
for achieving low turnaround times with large, intensive MTC applications.
However, HPC systems often lack a dynamic resource-provisioning
feature, are not ideal for task communication via the file system, and
have an I/O system that is not optimized for MTC-style applications. Hence,
additional software support is likely to be required to gain full benefit
from the HPC hardware.

\end{abstract}

\pagebreak

\setcounter{tocdepth}{2}
\tableofcontents

\pagebreak

\section{Introduction}

As computers have become more powerful,
both simulation and data-processing applications have become increasingly complex.
Simulations have increased in dimension (1D to 2D to 3D),
both in the equations being simulated (one equation to multiple equations in
one domain to multiple equations in multiple domain) and in the number of
time scales being studied simultaneously. Similarly, data-processing applications
have increased in terms of complexity of the analyses being run.
In both cases, the next natural step seems to be to increase the number
of such applications that fit into a meta-application.  This may involve 
adding another layer around the initial application, such as in optimization
or uncertainty quantification or in a parameter sweep.  Such applications
can be considered many-task computing (MTC) applications, since they
are assembled of a series of tasks, each of which may be a full application
or something simpler.
In a recent talk David Keyes identified
reasons why today's computational scientists want performance:
resolution, fidelity, dimension, artificial boundaries, parameter inversion,
optimal control, uncertainty quantification, and the statistics of ensembles~\cite{keyes}.
The last four of these can be addressed by MTC.

The term MTC first appeared in the literature in 2008, introduced to describe
a class of applications that did not fit neatly into the categories
of traditional high-performance computing (HPC) or high-throughput computing
(HTC)~\cite{FALKON-SC-08}.  Also in 2008,  a workshop titled
``Many-Task Computing on Grids and Supercomputers'' was held; this workshop
subsequently has been run at the SC08, SC09, SC10, and SC11 conferences.

As with traditional HPC, a defining aspect of MTC is the emphasis
on performing a large amount of computation in a timespan of days or even hours,
 in order to provide important results in a timely manner.
However, in contrast to traditional HPC applications, which tend to be a single program
(e.g., using MPI) run simultaneously on many nodes of a single cluster or
supercomputer, an MTC application is a set of many distinct tasks
with interdependencies, often viewed as a directed
graph of data dependencies.  In many cases, the data dependencies will
be files that are written to and read from a file system shared between
the compute resources; however, MTC does not exclude applications in which
tasks communicate in other manners.

For many applications, a graph of distinct tasks is a natural way to
conceptualize the computation and is often a natural way to build
the application, particularly if some tasks can be performed by
existing, standalone programs.
Structuring an application in this way also gives
increased flexibility.  For example, it allows tasks to be run on multiple
sites simultaneously; it simplifies failure recovery and allows the application
to continue when nodes fail, if tasks write their
results to disk as they finish; and it permits the application to be tested and
run on varying numbers of nodes without any rewriting or modification.

MTC applications can greatly benefit from being run on high-end HPC systems
such as Blue Waters, and candidate applications for HPC---those that require
high-performance hardware and timely results---may benefit from incorporating
ideas from MTC into their design.
The hardware of HPC systems such as Blue Waters, with its high degree of
parallelism and high-performance communication networks, is well suited
to achieve low turnaround times for large-scale, intensive MTC applications.
As we discuss in detail in this report, however, many MTC applications may not
be viable on HPC systems such as Blue Waters
without hardware and software systems support for
specific features.  For example, HPC systems often lack a fine-grained dynamic resource
provisioning feature and have I/O systems that are not optimized for
MTC-style applications.
MTC applications also generally require a node operating system (OS)
that supports full POSIX fork() and exec() semantics, in order that worker-node provisioning agents can execute the widest possible range of arbitrary application programs.
This report discusses problems presented by
MTC applications as a whole as well as by specific design patterns commonly
present in MTC applications. Many of these problems can be solved by using
appropriate middleware, but others may place additional requirements on
the underlying Blue Waters hardware and software environment.

\begin{figure}[h]
    \center
    \includegraphics[width=12cm]{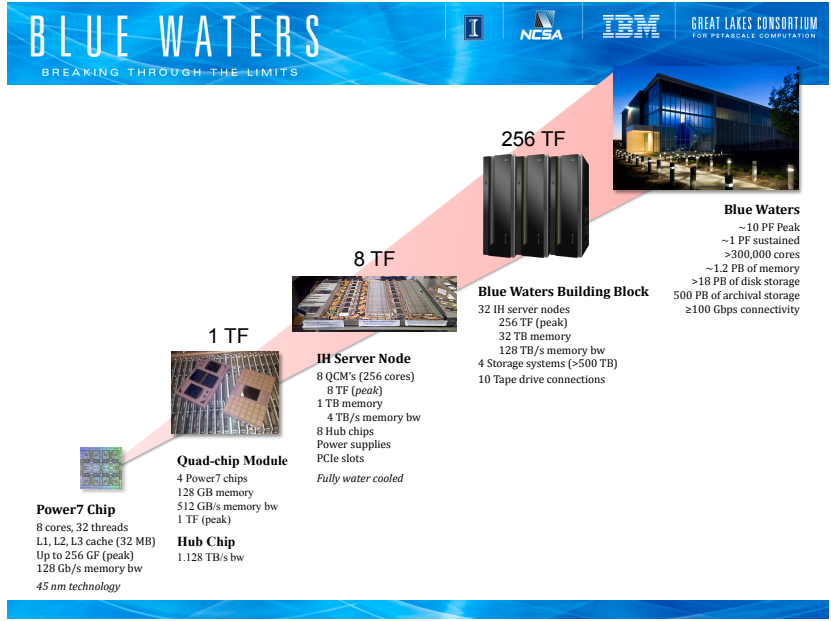}
    \caption {Processing hierarchy of the initial Blue Waters system.}
    \label{BWhierarchy}
\end{figure}

The initial design of Blue Waters was an IBM Power7-based system, with multiple levels
of hierarchy, going from cores that can run multiple threads through chips,
nodes, supernodes (drawers), and multirack building blocks, up to the full system,
all network connected, with different types of connections at different levels (see Figure~\ref{BWhierarchy}).
It was to have a systemwide, shared global file system (running GPFS) that would be embedded and distributed within
the network.  The global file system was to have integrated hierarchical storage management via HPFS for archival data retention.
Blue Waters was going to run a full Linux kernel on each 32-core compute node. Our understanding was that the nodes would boot off of the global shared file system and would have only RAM-disk for limited, node-local file storage. Nodes were to have approximate 4~GB total RAM available per core.  IBM and NCSA recently announced, however, that this initial version of Blue Waters will not be built~\cite{bw-Aug2011-announcement}. Instead, NCSA is planning a new design for the Blue Waters systems, but no details are currently public.

\subsection{Motivation: Making More Things Easy}

Petascale scientific computing poses multiple challenges that
must be addressed by the deployment teams who intend to deliver these
exceptional resources to application users.  Porting and
optimizing tightly coupled applications on new machines will be a
time-consuming endeavor and may not succeed in all cases.  Worthy
applications that can quickly be promoted to petascale should be addressed first,
regardless of the technologies used.  Beyond petascale, systems
researchers have identified challenges in communication, fault
tolerance, and other areas that will make continued increases beyond
Blue Waters-class systems difficult.

MTC offers the ability for domain scientists and system providers to
rapidly develop and deploy applications that can gain near-peak
hardware performance because of the nature of the application and software
structure.  Although not every application may be structured as an MTC
application, many can. \emph{when found scientifically worthy of allocations on
petascale resources, these applications should be provided with the
most practical tools to do the job}.

Some of the challenges petascale applications face, and the reasons
MTC applications can meet these challenges, are as follows:

\begin{itemize}
  \item {\bf Expression of natural parallelism:} Many algorithms found
    in applications from a wide variety of scientific domains are
    naturally divisible into cleanly separated task executions.  In
    their simplest form, these tasks are constructed as collections of
    independent POSIX processes, each of which consumes and produces
    data files over POSIX interfaces.
  \item {\bf Rapid application development:} The familiar POSIX
    computing model offers several advantages in the development of
    many-task applications.  Developers do not need to learn a new API in
    the language.  Individual tasks may be programmed as sequential
    programs, without threads or other multiprogramming models.
    Communication over traditional file system interfaces allows the
    use of customary file management techniques and tools.
    Additionally, these applications may be debugged on workstations
    or other computers by using standard methods and software.
  \item {\bf Portability:} The ability to scale many-task applications
    from workstations up to petascale systems not only aids in
    development but also \emph{frees the science team to use multiple
    resources, multiplying the return on the the initial development
    effort}.  Resources may then be selected by availability, the
    presence of specific performance-boosting hardware (e.g., GPUs),
    or other reasons.  Portability also facilitates code sharing among groups
    with access to different computational infrastructures.  Notably,
    this portability enables the use of grid resources such as the
    Open Science Grid.
  \item {\bf Fault tolerance:} The partitioning of application
    procedures into individual processes with well-defined input and
    output parameters enables the use of robust fault handling
    mechanisms familiar in other settings (e.g., exit codes) and the
    development of simple fault response strategies (e.g., re-execution)
    as well as more complex techniques.  Historically, this benefit
    is based on experience with wide-area computing techniques.
    \emph{On systems like Blue Waters, these time-tested patterns and
    methodologies will help users get real scientific applications
    up and running quickly}, both during system shakeout and in the
    presence of ongoing faults.
\end{itemize}

\subsection{Overview}

In the remainder of this report, we discuss previous work with MTC applications
(\S\ref{sec:PrevWork}), the results of a survey of cyberinfrastructure providers
regarding MTC applications on their systems (\S\ref{sec:PGIapps}), a survey
of MTC applications (\S\ref{sec:apps}), a categorization of MTC
applications (\S\ref{sec:categories}), and the hardware and software needed to
support MTC applications (\S\ref{sec:support}). We present conclusions
and recommendations in
\S\ref{sec:conclusions}.

\section{Previous Work\label{sec:PrevWork}}

MTC applications have emerged as a result of the wide impact of
distributed-computing and grid-computing application development in recent
years.  Applications developed for these platforms are necessarily
loosely coupled; cooperating processors may be located in distributed
locations, connected by wide-area networks.  These use cases led to
the development of programming languages and runtime systems that
enabled users to run and manage ever more jobs at larger scales.
However, the scale was ultimately limited by the constraints involved
in shared access to resources, making relatively few (tens of) processors
available to an individual at a time.

\subsection{System Software Support}

Porting these applications to massively parallel HPC systems enabled
individual users to run the same application at a very large scale,
increasing to the range of thousands the number of processors that can be applied to a given
application.  These applications require new
system support techniques to use the systems effectively.
First, the scheduler must be able to quickly allocate
processors for jobs without using existing heavyweight system
schedulers such as PBS~\cite{PBS_1998}.  An early solution, the
Falkon~\cite{falkon} scheduler, achieves high job submission rates by
allocating executor processes with the system scheduler, such as
Cobalt~\cite{Cobalt_WWW}, and scheduling tasks from the Falkon client
to distributed task dispatchers and finally to the task executors themselves.
This task scheduling mechanism bypasses the normal system scheduler
for individual user jobs, reducing a job execution to the time it takes for a short interprocess communications (IPC)
message exchange.

Second, appropriate data management and movement mechanisms must be
used to transfer data among HPC file services (e.g.,
GPFS~\cite{GPFS_2002}) and between intermediate system layers and user
processes~\cite{Collective_IO_2008}.  While MTC applications typically
access parallel file systems over POSIX interfaces, they cannot
directly benefit from parallel I/O
optimizations~\cite{ParallelIO_2008} as made available in
MPI-IO~\cite{MPI-IO_1999}.  Their use of the file system
typically appears as many small,
uncoordinated accesses to the file system, resulting in poor
performance.  However, applications patterns may be
observed and categorized~\cite{wozniak09petascalestorage} and then exploited by appropriate
software~\cite{Collective_IO_2008}.
More generally, aggressive caching may be used by
distributing data items across caches on the compute
sites~\cite{DataDiffusion_2008} or by employing a distributed
hash table~\cite{CMPI_2010}.

\subsection{Novel Infrastructures and Portability}

New infrastructure such as compute clouds, installed at commercial
data centers and research institutions, has additionally motivated
the use of many-task methodologies.  Compute clouds feature commodity
hardware components organized and managed in a highly economical,
scalable manner and emphasize flexibility through operating system
virtualization and on-demand resource
allocation~\cite{CloudOverview_2008}.  This type of infrastructure is
typically not associated with high-performance networks or other
features found in HPC installations, making it a natural target for
MTC-oriented applications such as
workflows~\cite{CloudWorkflows_2008}.  MTC applications are
additionally often compatible with opportunistic computing systems
such as Condor~\cite{Condor_Experience_2004} and grids such as the
Open Science Grid~\cite{OSG_2007}.

MTC applications offer this extreme portability by relying on widely
portable languages and system interfaces.  Typical use cases employ
POSIX-related shells and other high-level, widely available languages
such as Java and Python.  MTC domain-specific languages such as Swift~\cite{swift-ieee09,Swift_2011}
and Pegasus~\cite{Pegasus_2005} are in turn developed with these tools.  System interfaces
used by these systems are limited to the widely available POSIX-like
calls made available by these high-level languages.  If an application
can benefit from an infrastructure-specific
optimization, it must be made available by the MTC language and
runtime system; this is an active area of research.

\subsection{The MTC Community}

Current research, development, and production computing in MTC are
performed by a broad community of researchers, institutions, and user
groups.  MTC researchers are typically involved primarily in
traditional distributed or high-performance computing and contribute
to the MTC knowledge base tangentially.  Similarly, application groups
focus primarily on the specifics of their domain but contribute to the
model by producing problems (both practical and conceptual) to be
addressed by MTC systems, research, and development.  The
Many-Task Computing on Grids and Supercomputers (MTAGS) workshop
at SC acts as one focus where these groups interact.

\subsection{Task-Oriented Exploration and Problem Solving}

A decade ago or earlier, it was recognized that applications composed
of large numbers of tasks may be used as an driver for numerical
experiments that may be combined into an aggregate
method~\cite{Nimrod_G_2000}.  In particular, the following algorithm
paradigms are well suited for MTC:

\begin{itemize}
\item {\bf Optimization:} the process of exploring a parameter space
  to find extreme results.  This model consists of the creation of
  many experiments that provide sparse information about the space; a
  higher-level method is applied to solve the optimization problem,
  possibly through the creation of additional experiments.  MTC
  implementations treat individual experiments as tasks and use a
  higher-level program to make use of the results as a whole.
\item {\bf Data Analysis:} the concept of extracting aggregate or
  statistical information from existing data.  Implementations are
  often structured to gain high I/O rates relative to computation,
  possibly on different data storage sites.  MTC implementations
  provide a natural distribution of tasks and a model for generating
  useful final results.
\item {\bf Monte Carlo (MC):} the exploration of a system by
  performing random experiments within it, followed by an integration
  of results.  As in optimization, MTC implementations can be used to
  rapidly schedule randomly parameterized tasks and integrate results.
\item {\bf Uncertainty Quantification (UQ):} the determination of the
  quality of a result.  Computational results may be evaluated
  for sensitivity to perturbations in the input or numerical method
  used.  MTC investigations into UQ may be structured by integrating
  results from batches of individual task executions, formulated as
  a Monte Carlo investigation or other method.
\end{itemize}

\section{Production Grid and HPC Systems Survey\label{sec:PGIapps}}

We have asked a number of infrastructure providers about MTC applications, and found the following.

TeraGrid providers responded with two applications: work on hurrican ensembles from NOAA and work on an ensemble Kalman filter inverse problem for oil reservoir simulation that has tasks distributed over Abe, Queen Bee, and Ranger~\cite{yaakoub}.

Regarding the NOAA application, Bill Barth at TACC reported:

``We had, over the last two hurricane seasons, teams from NOAA doing ensemble weather forecasting for hurricane track prediction. These simulations were done both at the global and regional level using FIM and WRF, respectively. In these simulations, multiple runs of each case are simulated with slightly different initial conditions and incorporating the latest data from aircraft, satellites, buoys, and weather stations, and the results are averaged in clever ways to give a prediction of the hurricane path. The results of these simulations turned out to be much better than the methods NOAA was using at the time.

``Obviously, predicting the path of a hurricane directly can save both lives a dollars. The better the prediction, the few people who have to evacuate, and the more accurate the evacuation orders can be.''

Open Science Grid providers mentioned two applications: the Large Hadron Collider~\cite{CMS} and the Laser Interferometer Grravitational Wave Observatory~\cite{ligo}.  Both have been well studied and well characterized in previous publications.

From the Department of Energy, David Skinner at LBL stated:
``Most of the parameter sweeps I have been involved with in QCD and
chemistry have had MPI codes underneath. For example in quantum Monte
Carlo one often allows one task (or set of tasks) to quickly cancel
and reschedule work running on other tasks which results in rapid
pruning and growth of who is doing what. It's not deterministic but
rather like a workflow.''

David also discussed another case:
``Replica exchange MD is another case where an ensemble of parallel MD
runs rapidly communicate small pieces of information between each
other. It requires a parallel computer but there are two scales to the
level of interconnection.''

Katherine Riley at the Argonne Leadership Computing Facility highlighted the work of David Baker
at the University of Washington as ``a great example of a parameter sweep/ensemble set.''  His
laboratory focuses on the prediction and design of protein structures
and protein-protein interactions.  Baker's group created the Rosetta application~\cite{rosetta},
which has a BOINC-based Rosetta@home version~\cite{rosetta@home} and supports a server
called Robetta~\cite{robetta}.  Katherine also mentioned the work of Benoit Roux and Andrew Binkowski, who use the DOCK application in a manner similar to what we describe in \S\ref{sec:app:dock}.

Katherine described as another promising MTC application area the materials science research in `rational materials design led by Larry Curtis and Jeff Greeley.
Theory-aided design of novel materials is a growing area of research that has the potential to revolutionize the materials discovery process. Until recently, however, the ability to characterize many materials has been hampered by the lack of computer resources and by the difficulty for smaller organizations to harness large amounts of distributed resources and novel petascale systems. 
That situation is beginning to change with the introduction of petascale computers that allow for the rapid computational characterization of many candidate materials. For example, in catalysis studies it is possible to characterize 1,000 candidate surface compositions within a few hours on the Argonne Blue Gene/P machine. Moreover, the development of new density functional codes is enabling scientists to run more accurate computations in parallel on thousands of processors on large cluster, grid, petascale, and cloud systems. One such approach is being pioneered by Jens Norskov and his colleagues~\cite{Greeley2006,Greeley2009}, who are collaborating with Argonne's Center for Nanoscale Materials and Mathematics and Computer Science (MCS) Division to develop these codes for use on the Blue Gene.
Orchestrating the large numbers of computations demanded by the rational design process, whether on petascale computers or on other platforms such as scientific clouds, is a clear application for the MTC programming model.

Speculating on applications of MTC in finer-grained mathematical algorithms, Todd Munson of Argonne's MCS Division described five scenarios in which mathematical applications will require MTC execution patterns as they expand to petascale and exascale computing levels. These applications are summarized below, in order
of their ``shovel readiness.'' The application descriptions were edited for inclusion here from a private communication~\cite{Munson2010}.

\begin{description}
\item Sequential Monte Carlo with Reweighting for Climate Model Assessment

Purpose: uncertainty quantification
 
 Algorithm:
      Generate an ensemble of initial conditions and weights;
      run a climate simulation on each element of the ensemble
         in parallel, where each climate simulation runs on 128--1024 processors;
      analyze the results to compute uncertainty information and recompute weights;
      rerun the climate simulations with the new weights.

An interesting point here is that the analysis phase can
sometimes be run in parallel with the climate simulations.  It does
not always need the full set of results to compute the new weights,
limiting the amount of synchronization required.  Moreover, a
weight can be zero, in which case the corresponding element of the ensemble
no longer needs to be run and can be removed from the queue.
(If it is currently running, it can be stopped.)  It is possible that
the weights may provide priorities
and indicate the order in which to compute the elements of the
ensemble (i.e., one may need a priority queue where the weights
can be adjusted).

In this application, the number of processors required should
be constant and known ahead of time.

\item Uncertainty Quantification Using the Adjoint Method

Purpose: uncertainty quantification

  Algorithm:
    Run a PDE simulation forward in time on, say, 1024 processors 
      (the forward simulation checkpoints at regular intervals);
    run an adjoint simulation backward in time on, say, 1024*1024
      processors.  The adjoint simulation runs the
      forward PDE starting from one of the checkpoints to gather
      required information. Then the adjoint computation
      is rerun backwards.

An interesting resource utilization characteristic here is the ramp-up: one needs a relatively
small number of processors initially, but then more processors later.

\item Optimal Design with Integer Variables Using Branch and Bound

  Purpose: parameter estimation and design for partial differential equations
  
  Algorithm:
    Generate a set of subtrees using strong branching:
      for each integer variable solve two independent optimization problems 
        (each problem is a PDE-constrained optimization problem that runs on
          say 1024 processors).
    Once the subtrees are generated, run branch and bound on each
       of them independently.
    Each independent branch and bound on the subtree solves, for example, 1024
       optimization problems with PDE constraints, each of which requires
       1024 processors.

An interesting MTC characteristic for this algorithm is that some coordination is required in
the branch-and-bound process to prune subtrees and so forth.  Once the PDE-constrained
optimization problems are communicated, the additional
communication required for the branch-and-bound procedure is minimal.

\item Derivative-Free, Least-Squares Parameter Estimation

  Purpose: estimate parameters for a nuclear physics application
  
  Algorithm:
    Generate a prioritized list of trial points;
    run the trial points in parallel
      (each run with a trial point requires the evaluation of say 1024
        independent computations that can take vastly different
        amounts of time to complete);
   gather the results, update the priorities for the trial points, and add
      new trial points.

The interesting feature here is that running the trial points, 
updating the priorities, and adding new trial points can run in tandem.
Also, the 1024 independent calculations for each trial point have
priorities.  By observing the output from some of the initial calculations,
we can sometimes say immediately that the point is useless and that 
one should stop the rest of the calculations and pick the next trial point.

\item Hierarchical, Asynchronous Dynamic Programming

   Purpose: solve dynamic programming problems in parallel
   
   Algorithm:
     Solve a large set of optimization problems;
     gather the results to compute a new functional approximation;
     resolve a set of optimization problems.

The interesting MTC aspect here is that the data transfers are small.  Under
certain assumptions, solving the optimization problems can be
done simultaneously with updating the functional approximation.
A hierarchy also would need to be exploited.  This
is the least understood of the five applications and will need considerable mathematical analysis to assess its feasibility.

\end{description}

\section{MTC Applications Details\label{sec:apps}}
This section discusses the details for a set of MTC applications, describing 12
applications, their science domain, their history, and their computational characteristics.
Included is a discussion of how they were programmed, what infrastructure they use, what processing
paradigm they use, and what  their data and computing requirements are.  Many of these
applications have been previously
discussed, either individually as noted below or in summary~\cite{swift-ieee09}.

\subsection{AstroPortal}

AstroPortal~\cite{raicu06astroportal}
is an application that provides astronomers with the ability to
dynamically combine astronomical data sets to create new, ``stacked'' composite images.
Often, by combining images of different wavelengths or images  taken at different times, one
can detect astronomical objects that may be too faint or
indistinguishable from noise in a single image only.

\subsubsection{Science Domain} Astronomy

\subsubsection{History}
AstroPortal was developed from 2005-2007 to provide a service to
astronomers and also to investigate the dynamic analysis on-demand of large data
sets (in this case the Sloan Digital Sky Survey).

\subsubsection{Computational Characterization}

\begin {itemize}
\item {\em Language:} Java
\item {\em Infrastructure:} The system uses SDSS data, accessed from a GPFS file system.
A SDSS database is used to locate objects.  Globus Toolkit 4 services are used for
transferring data and launching jobs.
\item {\em Processing Paradigm:} A master-worker paradigm is used.  Each stacking is data-parallel.
\item {\em Data:} The size of the data set used in the development of AstroPortal (SDSS DR4)
is approximately 3 terabytes when compressed.  However, the amount of this image
data that is actually used in a given stacking job varies.  The data set
is divided into images of 2048 $\times$ 1489 pixels of approximately 2 MB
each.  The number of images that need to be stacked for a typical query invocation
is uncertain, but we can estimate the data size would be in the range of 5 to several
hundred megabytes, depending on the region of the sky.

\item {\em Processing Time:} The amount of processing required for each stacking is small.
Less than a second per stacking was required on 2006-era commodity hardware .

\end{itemize}

\subsection{PTMap}

Many biological processes are controlled by post-translational
modifications (PTMs) of proteins; such
PTMs are studied in order to understand the mechanisms of cell regulation.
PTMap~\cite{ptmap09} is used for mapping sites of PTMs
using mass spectrometry data and databases of protein sites.
Commonly, the algorithm will be invoked on all pairs
of input data.  Results are combined, selected by quality, and
reprocessed until high-quality results are obtained.

\subsubsection{Science Domain} Biochemistry

\subsubsection{History}

PTMap originated from a group at the Department of Biochemistry and
Pharmacology, University of Texas Southwestern Medical Center. As
members of the team moved to the Ben May Department of Cancer Research
at the University of Chicago, it is now a joint effort of Ben May
Department of Cancer Research and the Computation Institute. The
Cancer Department team is in charge of PTMap code development and
maintenance, while the Computation Institute group parallelizes the
code.

\subsubsection{Computational Characterization}
\begin {itemize}
\item {\em Language:} C++
\item {\em Infrastructure:} Cluster computing
\item {\em Processing Paradigm:} All-Pairs: Each spectroscopy given
  file is compared to each given protein sequence (FASTA) file.
  Further processing is derived from these results.
\item {\em Data:} Overall, 1.1~TB of data is read from the file system.
  For each protein, approximately 120~MB must be read. Each pair of
  spectroscopy datasets requires approximately 150~KB to be read.  The
  intermediate files, used for communication between tasks, are of the
  order of 100~KB per task.  The final output is very small.
\item {\em Processing Time:} Each pair requires 5--10 minutes on IBM BG/P, which has a 850-MHz quad-core CPU.  
Typical use case of the program may involve 50,000 or more
  tasks~\cite{wozniak09petascalestorage}.
\end{itemize}

\subsection{OOPS: Protein Structure Prediction}

The Open Protein Simulator (OOPS) builds on the Protein Library (PL).
Both have been developed at the University of Chicago.
OOPS is multipurpose and allows extensions to perform various simulation tasks
relevant for life scientists, such as protein folding or protein structure prediction.

\subsubsection{Science Domain} Biochemistry

\subsubsection{History}
OOPS research can be traced back to 2006, when it started
as the Open Protein Simulator, which was
created by a group of chemical and biological scientists. In 2008, it became
a joint effort of Department of Chemistry, the University of Chicago and the Computation
Institute. 

\subsubsection{Computational Characterization}
One situation where OOPS has been used in an MTC context is for protein structure prediction
using Monte Carlo--based simulated annealing (MCSA), a technique that
requires many randomized,
independent computations~\cite{hocky-oops-09}.  This is the
use case discussed here.
\begin {itemize}
\item {\em Language:} C++ for application code, Swift for coordination
\item {\em Infrastructure:} Swift interpreter and a compatible 
        task dispatcher deployed on the computational infrastructure;  
        shared file system.
\item {\em Processing Paradigm:} MTC, communicating via shared file system
\item {\em Data:} \begin{itemize}
    \item   Input: Common input data of 27~MB.  Per protein datasets (shared between iterations
                    of each task in the same proteins) in several files of sizes $\sim$1~KB
    \item Output: $\sim$1~MB (verbose mode), $\sim$1~KB (regular mode) per task. 
    \end{itemize}
\item {\em Processing Time:} 0.5 to 3 CPU-hours per MCSA task; approximately 1000 CPU-hours overall for a typical protein on a $\sim$2.33-GHz x86 CPU.

\end{itemize}

\subsection{DOCK\label{sec:app:dock}}

The DOCK6 molecular dynamics application identifies the low-energy binding modes of a small molecule (ligand) within the active site of a macromolecule (receptor). A compound acts as a drug if it inhibits the function of the receptor it binds to.
DOCK6 is used for the following purposes:
\begin{itemize}
\item Predict binding modes of small molecule-protein complexes
\item Search databases of ligands for compounds that inhibit enzyme activity
\item Search databases of ligands for compounds that bind a particular protein
\item Search databases of ligands for compounds that bind nucleic acid targets
\item Examine possible binding orientations of protein-protein and protein-DNA complexes
\item Help guide synthetic efforts by examining small molecules that are computationally derived
\end{itemize}

\subsubsection{Science Domain} Bioinformatics

\subsubsection{History}
The DOCK application family can be traced back to the 1980s, with a variety of versions, including DOCK3, DOCK4, DOCK5 and DOCK6. The DOCK program has been executed in parallel both in an MPI implementation and in a grid environment. As stated in~\cite{FALKON-SC-08}, DOCK6 has scaled up to 128,000 CPU cores on BG/P with the scheduling support of Falkon.

\subsubsection{Computational Characterization}
\begin {itemize}
\item {\em Language:} C++
\item {\em Infrastructure:} Swift interpreter and a compatible 
        task dispatcher deployed on the computational infrastructure.  
        Shared file system.
\item {\em Processing Paradigm:}
  \begin{itemize}
  \item HTC Mode: Each DOCK6 run is completely independent from others
  \item HPC Mode: (massively parallel) An MPI version of the master-worker model is implemented for DOCK6.
  \end{itemize}

\item {\em Data:}
  \begin{itemize}
  \item Input: 1 ligand file $\sim$10~KB, 1 grid.nrg file $\sim$10~MB, 1 dock.in file $\sim$1~KB,
    1 selected\_spheres.sph file $\sim$1~KB, 1 vdw\_AMBER\_parm99.defn file $\sim$10~KB,
    1 flex.defn file $\sim$1~KB, 1 flex\_drive.tbl file $\sim$1~KB.
  \item Output:1 scored ligand file $\sim$10~KB, 1 standard output file $\sim$1~KB.
  \end{itemize}

\item {\em Processing Time:} Processing time varies from seconds to hours on BG/L, with an average of 713$\pm$560 seconds~\cite{FALKON-SC-08}.
\end{itemize}

In HPC mode, the performance of DOCK6 starts to decrease significantly at the scale of 16,483 cores on BG/L~\cite{IBM-DOCK-08}, due to
    \begin{enumerate}
    \item Scheduling capability of a single master node
    \item Data-processing capability of a single master node, as all input files are read in by the master node then randomly distributed to slave nodes
    \item Unpredicted load balancing caused by random data distribution
    \item I/O capability between compute nodes to GPFS
    \end{enumerate}
To improve the load balancing, we could sort the input files according to the running time, as the running time could be predicted by ``the greatest number of rotatable bonds'' and ``the number of atoms per ligand''.
    
In HTC mode, there is linear scalability up to 16,384 cores~\cite{IBM-DOCK-08}. Work with Falkon showed a sustained utilization of 99.6\% in the first 5700 seconds out of 7200 second run on 128,000 cores.~\cite{FALKON-SC-08}

\subsection{Montage\label{sec:montage}}

The purpose of Montage (\url{http://montage.ipac.caltech.edu/})~\cite{montage1,montage2} is to build astronomic image mosaics, while preserving image accuracy.
That is, the amount of energy in the input images is conserved and the position of the energy is preserved.

\subsubsection{Science Domain} Astronomy

\subsubsection{History}

Montage development began in 2002, with the first production release in 2003 (v. 1.7).  The current version (v 3.0) was released in 2007.  Version 3.2b6 is the current release.

\subsubsection{Computational Characterization}

\begin {itemize}
\item {\em Language:} C for application code, MPI or Pegasus for infrastructure
\item {\em Infrastructure:} shared file system for MPI, with Pegasus handling file transfers
\item {\em Processing Paradigm:} Many-task computing, communicating via shared file system
\item {\em Data:} A benchmark problem is
generating a mosaic of 2MASS data from a 6 degree $\times$ 6 degree
region around M16.  Construction of this mosaic requires 1,254 2MASS
images as input, each having about 0.5 megapixels, for a total of
about 657 megapixels input (or about 5 GB, with 64 bits per pixel
double-precision floating-point data).  The output is a 3.7 GB FITS
file with a 21,600-pixel $\times$ 21600 pixel data segment and 64
bits per pixel double-precision floating-point data.  Note that the
output data size is a little smaller than the input data size because
of some overlap  between neighboring input images.
\item {\em Processing Time:} For the benchmark problem above, about 280 minutes on a single 1.5 GHz core, and about 20--30 minutes on 64 such cores.
\end {itemize}

Montage takes a number of image files as input and builds an output file (possibly tiled) as output.  The output image can be almost as large as the input images (minus the overlaps in the inputs), or it can be smaller if the resolution/projection of the output is significantly different from that of the input.

Typical steps in Montage:
\begin{itemize}
\item Reprojection of input images to a common spatial scale, coordinate system,
and WCS projection (multiple tasks can be done in parallel)
\item Modeling of background radiation in images to achieve common flux scales
and background levels by minimizing the interimage differences (initially multiple tasks can be done in parallel, followed by a sequence of tasks that must be done in order)
\item Rectification of images to a common flux scale and background level (multiple tasks can be done in parallel)
\item Co-addition of reprojected, background-corrected images into a final mosaic (can be done in parallel in MPI but is single task)
\end{itemize}

\subsection{Social Learning Strategies}

Computer simulations can be used to provide insight into the role social
learning plays in evolution, and human behavior.
One simulation places two different kinds of learning agents---social
and asocial---on a two-dimensional grid, with reproduction, environmental change,
movement, and learning playing a role in the simulation~\cite{rendell-sociallearning-09}.

Another application, a tournament~\cite{rendell-sociallearning-10}, involves
competitions between a number of autonomous agents submitted by a number of
different
participating groups.  Each match in the tournament
is a contrived game, where each agent follows its own strategy to maximize
its payoff according to the rules of the game.  The choices each agent must
make at each step are designed to be somewhat analogous to those a person
must make when learning and performing a  new task.

In both the simulation and the tournament, the relative success and failure of
different learning strategies were used to derive insight into
the role and importance of social learning to humans.

\subsubsection{Science Domain} Psychology, Evolution

\subsubsection{History}
Similar tournaments focusing on simulated social
behavior were initially organized in the 1980s by Robert Axelrod~\cite{rendell-sociallearning-10}.

\subsubsection{Computational characterization}

\begin {itemize}
\item {\em Language:} Matlab/GNU Octave
\item {\em Infrastructure:}  A desktop computer was used for the simulation.  The UK National Grid Service (NGS) was used for the tournament.
\item {\em Processing Paradigm:} All-Pairs: each pair of strategies was tested
    against each other, with multiple randomized simulations performed for
    each pair.
\item {\em Data:} Minimal size.  The tournament required strategy definitions
            as input.  The output is statistics about the success
            of different strategies.
\item {\em Processing Time:} In both cases, 5--20 minutes per simulation
            instance on a single $\sim$2.5-GHz core was required.
    \begin{itemize}
        \item {\em Simulation:}  Each simulation involved 2,000 rounds,
        simulating several thousand agents.  There were 20 instances
        of a simulation per parameter combination.  The simulations were
        run in batches of about 600 parameter combinations to investigate
        different factors; each batch took 2--3 days on a single-core machine~\cite{rendell-email-10}.
        \item {\em Tournament:}
        The first stage of the tournament involved pairwise competitions among
        104 strategies, with 10,000 rounds in each competition.
        The second stage involved the 10 best strategies,
        where they all competed in the same simulation rather than in pairwise
        matches.  Approximately 65,000 CPU-hours were used on
        the NGS for the entire tournament~\cite{rendell-email-10}.
    \end{itemize}
\end {itemize}

\subsection{BLAST}

The Basic Local Alignment Search Tool (BLAST) finds regions of local similarity between biological sequences. The program compares nucleotide or protein sequences to sequence databases and calculates the statistical significance of matches. BLAST can be used to infer functional and evolutionary relationships between sequences as well as help identify members of gene families.

A number of varieties of BLAST exist:
\begin{itemize}
\item {nucleotide blast:} Search a nucleotide database using a nucleotide  query
\item {protein blast:} Search protein database using a protein  query
\item {blastx:} Search protein database using a translated nucleotide query
\item {tblastn:} Search translated nucleotide database using a protein  query
\item {tblastx:} Search translated nucleotide database using a translated nucleotide query
\end{itemize}

\subsubsection{Science Domain} Bioinfomatics

\subsubsection{History}
The BLAST program was designed by Eugene Myers, Stephen Altschul,
Warren Gish, David J. Lipman, and Webb Miller at the NIH and was
published in J. Mol. Biol. in 1990~\cite{blast-1990}.
BLAST is one of the most widely used bioinformatics
programs, because it addresses the fundamental problem of sequence
alignment, with an emphasis on speed.
This emphasis on speed is vital
for making the algorithm practical on the huge genome databases
currently available, although later algorithms can be even
faster. Before fast algorithms such as BLAST and FASTA were developed,
doing database searches for protein or nucleic sequences
by using a full alignment procedure like
Smith-Waterman was very time consuming. BLAST
cannot guarantee the optimal alignments of the query and database
sequences as Smith-Waterman does, but the results have proven to
be sufficiently accurate to prompt widespread adoption of BLAST by
scientists.

\subsubsection{Computational characterization}
\begin {itemize}
\item {\em Language:} C++
\item {\em Infrastructure:}  The Swift version requires a Swift interpreter and a compatible 
        task dispatcher deployed on the computational infrastructure.  
        Shared file system.
        
\item {\em Processing Paradigm:}
  \begin{itemize}
  \item HTC Mode: Each BLAST run is completely independent from others.
  \item HPC Mode: (Massively Parallel) An MPI version of BLAST is implemented by Wu Feng, director of the Synergy Lab at Virginia Tech.
  \end{itemize}

\item {\em Data:}
  \begin{itemize}
  \item Input: 1 database file $\sim$1~GB (a common database file is 6 GB), 1 query string file $\sim$1~KB
  \item Output: 1 text output file $\sim$1~KB
  \end{itemize}

\item {\em Processing Time:} A simple query may take $\sim$1 minute on BG/P. If multiple queries are wrapped in one transaction, the running time is longer.
\end{itemize}

In  HPC mode, mpiBLAST showed a 93\% efficiency performance on 32,768 cores on IBM BG/P at Argonne~\cite{mpiBLAST-SC-08}.

In  HTC mode, no productive measurements have been published. The machine utilization of current HTC implementations of BLAST is unacceptably low on supercomputers,
  because of load-balancing problems and due to BLAST's need to share the typically large search database in-RAM between cooperating SMP cores.~\cite{mpiBLAST-SC-08}

In summary, BLAST is a perfect example of an application for which MTC processing on petascale systems can be highly desirable and scientifically important and which is best accomplished as a hybrid application of many independent executions of tightly coupled mpiBLAST runs. This approach provides the workflow flexibility of restart and variable-sized work units while leveraging the efficient data sharing of tightly coupled MPI application kernels.

\subsection{CIM-EARTH}

CIM-EARTH is a collaborative, multi-institutional project to design a large-scale integrated modeling framework as a tool for decision makers in climate and energy policy. CIM-EARTH is intended to enhance economic detail and computational capabilities in climate change policy models and to create and support a broad interdisciplinary and international community of researchers and policymakers.

\subsubsection{Science Domain} Economics, Earth Systems, Climate

\subsubsection{History}
The CIM-EARTH project originated as a collaboration of researchers from University of Chicago, Argonne National Laboratory, and the Hoover Institute. Discussion about CIM-EARTH started around 2008. In 2009, a $\sim$50,000 CPU-hour run of the v0.1 CIM-EARTH model, expressed in the AMPL mathematical programming language, was performed on the TeraGrid Ranger supercomputer at TACC and on several OSG sites, including Firefly at the University of Nebraska-Lincoln and TeraPort at the University of Chicago.

The model is executed in large ensemble runs of 1,000 to 10,000 model executions with stochastic inputs, to perform uncertainty quantification.
A first v0.5 CIM-EARTH model coded as a native C++/FORTRAN app is under development, which will enable far greater portability of the model to an abundant set of petascale systems, including the BG/P, where we are currently unable to execute the binary AMPL application. 

\subsubsection{Computational characterization}
\begin {itemize}
\item {\em Language:} AMPL and TAO libraries for numerical programming and optimization, Swift for parallel processing.
\item {\em Infrastructure:} The workflow is written in Swift, execution is done  on clusters.  Requires data to be accessible on shared file system.
\item {\em Processing Paradigm:}
  \begin{itemize}
  \item HTC Mode: Each AMPL run is completely independent from others.
  \item HPC Mode: N/A
  \end{itemize}

\item {\em Data:}
  \begin{itemize}
  \item Input: 6 input files, $\sim$10~KB--100~KB each.
  \item Output: 6 output files, 2 of which are $\sim$10~MB size, 1 $\sim$100~KB, 2 $\sim$10~KB, with a $\sim$1~MB standard output.
  \end{itemize}

\item {\em Processing Time:} $\sim$10 minutes to $\sim$1 hour using $\sim$2.5-GHz cores.
\end{itemize}

Some experimental results have been published~\cite{CIM-EARTH-10}, but no performance results have been presented.

\subsection{SYNAPPS}

The purpose of SYNAPPS is to estimate the properties of a particular class of supernovae in such
a way that they match an observed spectrum. SYNAPPS uses the SYNOW spectrum synthesis fitter to
estimate the spectrum of a supernova using a model with over 50 input parameters.  Previously,
SYNOW was typically utilized in a human-supervised manner, with a human in the loop adjusting
the input
parameters until a reasonable fit was found, guided by experience and intuition rather than
any systematic approach.  SYNAPPS replaces the human with the APPSPACK optimization code to
iteratively explore the parameter space with many parallel instances of SYNOW~\cite{nugent-synapps-08}.

\subsubsection{Science Domain} Astronomy

\subsubsection{History}
The original version of SYNOW was released in 1995.
It is still being maintained, with an updated version 2.0 described in a
2007 paper~\cite{branch-synow-07}.  It has been used in a number of investigations
of supernovae.

SYNOW is in frequent use in the astronomy community.  A Google Scholar search for ``SYNOW supernova''
 returns 187 results, almost of all of which describe SYNOW being used for  astronomical research.

SYNAPPS was developed more recently and has been used to generate published
results since at least 2008.

\subsubsection{Computational Characterization}

\begin {itemize}
\item {\em Language:} SYNOW is written in FORTRAN. APPSPACK is written in C++/MPI.
\item {\em Infrastructure:} MPI
\item {\em Processing Paradigm:} Master/Worker.  Each worker is assigned a task, which is
            to calculate the ``figure of merit'' for a single point in the parameter space.
\item {\em Data:} \begin{itemize}
    \item Input: Observed spectrum (same for all tasks) of $\sim$60~KB, plus the model parameters for each task, which are $\sim$1~KB in size.
    \item Output: Figure of merit, which is a single number
    \end{itemize}
\item {\em Processing Time:} A few seconds for each task; less than two hours of wall-clock time
                total on a 96 Opteron processor cluster
\end {itemize}

\subsection{Deem's Database of Hypothetical Zeolite Structures}

Deem's database is a repository for zeolite structures (a kind of mineral)
that have been computationally predicted to have some reasonable
chance of existing in nature or being synthesizable~\cite{deem-2009}.  There are various applications
for such a database, such as identifying materials, searching for materials with
certain properties, or researching the properties of Zeolite structures.  Populating this
database has been a computationally expensive task, requiring an
extremely computationally expensive Monte Carlo search.

Constructing the database required over 3 million Condor jobs to be run,
each of which is a single
core task that takes between 10 and 60 minutes on an x86 processor.
There are no dependencies between jobs.  Various 
platforms were used to scavenge cycles for this application.
Over the 40-day period these 3 million jobs were run in, on average
200+ processors, which were used for a workflow involving three primary
applications.

\subsubsection{Science Domain} Physical Chemistry

\subsubsection{History}
Computationally predicting the structure of zeolites has been of interest since at least 1992,
when a Monte Carlo method for structure prediction was show to be valid in predicting real-world structures. 
Public and proprietary databases have grown in size since then; there
were 600,000 hypothetical zeolites known in 2006. The work on Deem's database has been ongoing since at least 2006, with large computing allocations on TeraGrid~\cite{deem-2009}.

\subsubsection{Computational Characterization}
An earlier paper describes some of the computational work done
in order to populate the database~\cite{walker-07}.
\begin {itemize}
\item {\em Language:} DAGMan for workflows
\item {\em Infrastructure:} Condor for task dispatch, MyCluster as middleware, running on
                        several TeraGrid resources
\item {\em Processing Paradigm:} HTC workflow, using a directed acyclic graph (DAG) to represent data dependencies
\item {\em Data:} \begin{itemize}
    \item Input: a set of parameters describing a region in the space of possible 
    zeolite structures
    \item Output: 3 million hypothetical zeolite structures, a few kilobytes each
    \end{itemize}
\item {\em Processing Time:} 2 million CPU-hours on $\sim$2.5-GHz cores.  Each task ran for a highly variable
            amount of time, from a minimum of few minutes to 3--4 hours for typical tasks.  It was possible for some simulations to run for an extremely long period of time, in which case they were terminated after 10 hours and not rerun.
\end {itemize}

\subsection{fMRI}

The fMRI application analyzes brain regions for response to experimental stimuli. A relational database of responses for a given subject may be queried for analysis, providing statistical connections to be made between MRI data and brain function. The fMRI script pulls records from the MRI database, performing statistical tests on each brain region using the statistical analysis language R, then writes the result.

\subsubsection{Science Domain} Neuroscience

\subsubsection{History}
The papers related to this research can be traced back to 1994~\cite{fMRI-1994, fMRI-1995}.

\subsubsection{Computational Characterization}
\begin {itemize}
\item {\em Language:} R for statistical computation, Swift for parallel processing
\item {\em Infrastructure:} The workflow is written in Swift, execution is done on TeraGrid. Requires  data to be accessible on a shared file system.
\item {\em Processing Paradigm:}
  \begin{itemize}
  \item HTC Mode: Each fMRI task is independent. Then a summary stage is applied to all previous results.
  \end{itemize}

\item {\em Data:}
  \begin{itemize}
  \item Input: 1 input file, 1~KB each
  \item Output: 1 output file, 2~KB each for first stage, 1 2-GB output file for the second stage
  \end{itemize}

\item {\em Processing Time:} $\sim$10 minutes to $\sim$1 hour on $\sim$2.5-GHz cores.
\end{itemize}

 In HTC mode, fMRI along with Swift has run on TACC's Ranger at the scale of 65,536 jobs in about 16 hours.

\subsection{Model SEED: Genome-scale Metabolic Models}

Genome-scale metabolic models are a valuable resource to understand
the genome, phenotypes, and behavior of the cells in an organism.  These metabolic models
are invaluable as they can be used to understand the function and importance
of different genes and can model the organism's behavior under different
conditions.  The models make testable predictions that can
be checked experimentally.  A key
component of each model is a set of gene-protein-reactions mappings, which
represent a theory about which metabolic reactions occur in the
organism's cells, which enzymes catalyze the reactions, and which genes
encode those enzymes.  In addition, the models 
require further information about the reactions that play a role in the
organism's metabolism and require a biomass objective function capturing
the molecules required for growth~\cite{henry-09}.

High-throughput sequencing of genomes has meant that the
pace at which genome sequences are assembled
exceeds the pace at which metabolic models can be manually constructed
using the genomes~\cite{henry-09}.  Manually reconstructing these models
can take 96 steps under at least one published protocol~\cite{henry-10}.

Model SEED (\url{http://blog.theseed.org/model_seed/}) is an online resource 
that automates much of the process
of constructing metabolic models from sequenced genomes.
It reconstructs a preliminary metabolic model
based on existing genome annotations, automatically filling in gaps in the
model.  This preliminary model can be checked manually, after which
Model SEED can automatically
optimize it through various analysis steps that
check the viability of the model, validate it against experimental data,
and perform optimizations to refine the set of reactions
included in the model~\cite{henry-10}.

The majority of the computation performed is in the form of
optimization problems.  The most computationally intensive
steps are those that attempt to add  to or remove reactions from the model, which
are formulated as complex mixed-integer linear
optimization problems (MILPs).  These require from one minute
to one day per organism to solve running on eight processors
in parallel~\cite{henry-09}.  Other steps, such
as simulations of an organism in different environments or with different genes removed, can be formulated as less computationally
intensive linear programs. However, these steps may require many more
problem instances to be solved.
For example, a simple growth simulation of an organism with a
set of genes removed
in a particular medium can be completed in around 20 milliseconds, but
a user might want to perform that simulation for millions or billions of
gene combinations in hundreds of different media per organism.  More
complex growth simulations, which check that predictions are thermodynamically
feasible, can take much longer: around 4 seconds per simulations~\cite{henry-09}.
All these computational steps may need to be performed
for hundreds or thousands of genomes, motivating the use of HPC.

\subsubsection{Science Domain} Bioinformatics

\subsubsection{History}
The model SEED has its origins in the Project to Annotate
1,000 Genomes, which was initiated by the Fellowship for Interpretation of Genomes
in December 2003.  The stated goal of this project was to produce and
make freely available high-quality annotations and metabolic models for
the first 1000 sequenced genomes~\cite{seed-manifesto-04}.

\subsubsection{Computational characterization}
\begin {itemize}
\item {\em Language:} C/C++
\item {\em Infrastructure:}~\cite{henry-email-10}
    \begin{itemize}
        \item Scheduling and distribution of work used for customized MPI code
        \item GLPK and CLP solvers used for linear optimization
        \item CBC and Scip solvers used for mixed integer linear optimization
        \item MySQL database used for data storage
    \end{itemize}

    \item {\em Processing Paradigm:}
        Master-worker and static scheduling approaches both used within
        MPI
    \item {\em Data:}
        Input and output files are in the range of 1--4~MB. Some
        stages require the input to be broadcast to all tasks; some stages
        have different per task input files.
        Task outputs are typically at most 2~MB~\cite{henry-09, henry-email-10}.
\item {\em Processing Time:}~\cite{henry-09}
    \begin{itemize}
    \item Up to 24 hours (on BG/P) per MILP task for automatic filling of gaps
            or generation of gaps
    \item 10--20 milliseconds for simple simulation tasks
    \item Minutes to hours for more complex simulation tasks
    \end{itemize}
\end{itemize}

Gene knockout simulation have scaled to 65,536 processors on Blue Gene/P
        {\em Intrepid}~\cite{henry-09}.

We note that as more high-quality options become available for various mathematical programming and optimization approaches, and these tools can be portably compiled across a broader set of system architecture and Linux/GCC/Libc variants, new opportunities will be made available for MTC applications across the newest and continually expanding set of petascale machines. In particular, the ability to run such tools on non-Intel architectures has been a limiting factor on petascale systems such as BG/P and will be limiting on Blue Waters as well, until such tools enter the open source toolkit.

\section{Categorizing MTC Applications\label{sec:categories}}

We present two methods for characterizing MTC applications.
In \S\ref{sec:apps:abstract}, we
discuss some general issues about the applications and the resources on which they run.
In \S\ref{section:patterns}, we discuss a set of patterns that are found in MTC applications.

\subsection{Abstract Issues of Applications\label{sec:apps:abstract}}

One method for analyzing MTC applications is to ask a set of questions about the application.
An issue that occurs during this process is defining ``the application.''  Is the application
the source code?  Or the algorithm?  Or the combination of the code and the general
system it is defined to run on?  What happens when an application has variants for
multiple systems?  Unfortunately, this issue dramatically complicates discussions
about the application.  Here, we discuss abstract issues about the application,
but we also discuss some issues about the resources used, and we focus
on a particular ``production'' use of the application, as defined in \S\ref{sec:apps}.

Some of the issues are properties of the application:

\begin{itemize}
\item What the type of task is involved: sequential or parallel (MPI) or both?
\item Is the communication at the start and end of the tasks (likely via files), or continuous within the tasks (via messages)?
\item How intense is the communication? (This leads to two types of parallelism, tightly coupled and loosely coupled, neither of which is rigorously defined.)
\item How many tasks does the application comprise?
\item Is the number of tasks defined at build time or runtime?  If it is defined at runtime, is it a function of the amount of input data or of the values of that data?
\item What is the shape of the graph of tasks? 
\begin{itemize}
\item Are any of the control-flow or data-flow patterns from \S\ref{section:patterns} present in the graph?
\item Is the graph divided up into distinct stages, with each stage dependent
on the previous?
\item Are there tasks that depend on the output of other tasks?  If so, how deep is the graph? That is, what is the length of the longest path from start to end?
\item Are there higher-level patterns in the graph, for example, MapReduce or AllPairs?
\item Does the graph have cycles in it? Is there a subgraph that is 
    iterated over and unfolded into a DAG at runtime?
\item How does the degree of parallelism (i.e., the width of the graph) change 
    as execution of the workflow proceeds?
\item Is the graph shape static or dynamic? That is, can tasks be added
    or removed from the graph during runtime?
\end{itemize}
\item How much computation and communication are there in the application?
\end{itemize}

Other issues are properties of the environment or resources on which the application is run:

\begin{itemize}
\item How long do tasks run? 
\item Is computation or communication the limiting factor in application performance?
\item Is the resource model static or dynamic? That is, can the set of resources that the application uses be changed during execution?
\end{itemize}

\subsection{Patterns\label{section:patterns}}

\begin{table}[t]
    \caption {MTC applications discussed in this report and corresponding patterns.  }
    \label{MTCAppsStats}
    \center
    { \footnotesize \renewcommand{\tabcolsep}{3pt}
    \begin{tabular}{|m{3.5cm}||>{\centering\arraybackslash}m{0.5cm}|>{\centering\arraybackslash}m{0.5cm}|>{\centering\arraybackslash}m{0.5cm}|>{\centering\arraybackslash}m{0.5cm}|>{\centering\arraybackslash}m{0.5cm}|>{\centering\arraybackslash}m{0.5cm}|>{\centering\arraybackslash}m{0.5cm}|>{\centering\arraybackslash}m{0.5cm}|>{\centering\arraybackslash}m{0.5cm}|>{\centering\arraybackslash}m{0.5cm}|}
            \hline
                        &\begin{sideways} Parameter Sweep  \end{sideways} & \begin{sideways}Data-intensive \end{sideways}& \begin{sideways}Scatter \end{sideways}  & \begin{sideways}Gather  \end{sideways} & \begin{sideways}Repartition \end{sideways}  & \begin{sideways}Iteration \end{sideways}	& \begin{sideways}Task Pruning \end{sideways} & \begin{sideways}Pipeline \end{sideways} & \begin{sideways} {\space Coordination of MPI Apps \space} \end{sideways}	& \begin{sideways}Variable Runtimes \end{sideways}\\
            \hline \hline
    AstroPortal	        &	            &{$\times$}	    &	        &	    &	            &	        &	        &	    &	                        &\\
            \hline
    PTMap	        &	            &{$\times$}	    &{$\times$} &{$\times$} &               &	        &	        &{$\times$} &	                        &\\
            \hline
    OOPS	        &{$\times$}	    &{$\times$}	    &{$\times$} &           &	            &{$\times$}	&   	        &{$\times$} &{$\times$}	                &{$\times$}\\
            \hline
    DOCK	        &	            &{$\times$}	    &{$\times$}	&{$\times$} &               &	        &	        &	    &{$\times$}	                &{$\times$}\\
            \hline
    Montage	        &	            &{$\times$}	    &           &{$\times$} &{$\times$}	    &	        &	        &{$\times$} &	                        &\\
            \hline
Social learning strategies &{$\times$}	    &	            &	        &	    &	            &	        &	        &	    &	                        &\\
            \hline
BLAST (MTC version)     &	            &{$\times$}	    & {$\times$}&           &	            &	        &	        &	    &{$\times$}	                &\\
            \hline
    CIM-EARTH	        &{$\times$}	    &	            &   	&	    &	            &	        &	        &           &	                        &\\
            \hline
    SYNAPPS	        &{$\times$} 	    &	            &   	&	    &	            &{$\times$}	&   	        &	    &	                        &\\
            \hline
    Deem's database       &{$\times$}	    &	            &   	&	    &	            &	        &	        &	    &	                        &{$\times$}\\
            \hline
    SEM	                &	            &	            &   	&	    &	            &	        &	        &	    &	                        &\\
            \hline
Metabolic Models (SEED)	&{$\times$}	    &	            &{$\times$} &           &	            &	        &{$\times$}	&	    &	                        &{$\times$}\\
            \hline
    \end{tabular}
    }
\end{table}

Summarizing the \S\ref{sec:apps} discussion of application
characteristics, Table~\ref{MTCAppsStats} shows that many applications
make use of computational and dataflow patterns.  These patterns,
known to the developer, may be used as a guide when developing MTC
middleware, enabling developers to gain access to toolkits optimized
for particular implementations.  In this report, we focus on
optimizations made possible by the advanced technologies available on
Blue Waters.  We outline some patterns here:

\begin{itemize}

\item Control Flow Patterns:

  \begin{itemize}
  \item Parameter sweep:  when the same code is
    run many times in parallel with only difference being input parameters.
    Randomized simulations can also be considered parameter sweeps,
    with the crucial parameter being the random seed.
  \item Iteration: when a section of the job is
    iterated a number of times, and the number is often unknown at the outset of the job, as it is
    determined by some loop completion criteria.
  \item Task pruning: when tasks are speculatively
    executed, and can be ``pruned'' dynamically before completion.
    Typically, this would occur when one task determines
    that the result of another concurrent task is no longer necessary.
    An example
    is a branch-and-bound algorithm, where branches are
    speculatively explored in parallel.
  \end{itemize}

\item Dataflow Patterns:

  \begin{itemize}
  \item Scatter: At a stage of the MTC computation, a single
    item of data must be broadcast to all subsequent tasks, or multicast
    to some number of subsequent tasks.  This
    could be the output of a previous task, or it could be a
    data file from the global file system (e.g., the sequence
    for a BLAST query). 
  \item Gather: A small number of tasks take as input the
    output from a large number of tasks.  Examples include a task
    that checks the results of previous tasks for a convergence
    criterion and a task that calculates summary statistics from the
    output of many tasks.
    \item Pipeline: A set of tasks operate on given data in
    sequence, with the output of one task becoming the input of the
    next.
  \item Data reuse: Bulk data is reused by task after task.
  \end{itemize}

\item Other Features:

  \begin{itemize}
  \item User toolkits with compositional flexibility:
  Many applications are structured as a set of tools (often, ``command line'' invocable) with moderately complex usage patterns, which benefit from a workflow language in which one can concisely and easily specify the execution patterns. The more declaratively these patterns of composition can be specified, the more work has been moved from the user to that of the workflow automation tool.
An excellent example of such applications is the Montage suite (\S\ref{sec:montage}), which gives astronomers a very general set of software modules that are efficiently and flexibly linked through loosely coupled file exchange.
  \item Coordination of MPI applications: Instead of sequential tasks,
    the unit of execution is a tightly coupled MPI application.
    Note that we contrast here applications of MPI in which the communicating processes conduct many message exchanges over their mutual lifetimes, from applications in which a communicating entity is a single task that gets its input, performs a process to completion, and either terminates or awaits a similar task. The former class we categorize as tightly coupled applications, while the latter are more loosely coupled and often exhibit MTC patterns.
    Again, Montage provides an excellent illustration of both these degrees of coupling. Some uses of MPI within Montage leverage repetitive tightly coupled message exchange, while others follow the input-process-output model and are well suited for MTC execution via file exchange. Enabling the loose coupling to be specified by a declarative functional workflow language, as we have done by specifying Montage workflows using Swift, illustrates the flexibility and scientific productivity afforded by the MTC model to manage the complex data dependency graphs that such applications entail.
  \item Variable runtimes: Concurrent tasks in the application
    can run for variable, possibly unpredictable lengths of time.
  \end{itemize}

\end{itemize}

\section{Support for MTC Applications\label{sec:support}}
This section enumerates a range of key challenges that arise when attempting
to run many-task applications on HPC systems. We first describe (\S\ref{sect:basicreq}) 
the basic requirements needed to run a typical many-task application.
We then discuss
(\S\ref{sect:middleware-design}) design choices of the requirements.
We conclude the section with a discussion
(\S\ref{sect:patternsupport}) of commonly occurring patterns of
data movement and communication, the challenges in efficiently supporting
them, and how we address those challenges with the design choices covered in
\S\ref{sect:middleware-design}.

\subsection{Basic Hardware and Software Environment}
\label{sect:basicreq}

We describe the requirements of MTC in three categories: hardware requirements, operating system requirements, and MTC middleware requirements.

\begin{itemize}
    \item Hardware Requirements:
      \begin{itemize}
	\item Local Storage: 
	  Local storage for compute nodes could be in the form of RAM disk, hard disk, or solid state disk. Local storage is used for local data caching and intermediate data storage. The data policy for local storage varies. For most supercomputers, the availability of local storage is synchronized with the allocation of compute resources, meaning that data on local storage is erased as the compute allocation is released.
	\item Network: 
	  Compute nodes require networks for communication and data collection. The network is usually a global network through which compute nodes can reach all their peers. Vendors have different network configurations: for example, some machines are built with specialized technologies such as InfiniBand, while others are built with commodity network technologies. Since a network is required for Message Passing, it is available on every high-end machine.
	\item Persistent Storage:
	  A global shared file system is required to store the input and output of MTC applications. Ideally, the shared file system can handle huge amounts of concurrent read/write operations.
      \end{itemize}
    \item OS Requirements:
      \begin{itemize}
        \item Full OS kernel:
	  Many scientific applications have been written assuming
          that various features are provided by the operating system on the
          compute node.
          Many applications will make use of some subset of
          system calls and functions from one of the POSIX standards.
          Others will also use OS-specific extensions, such as
          Linux-specific systems calls or functions from the glibc library.
	\item fork()/exec() support:
	  The fork()/exec() family of system calls is necessary for MTC middleware to run on supercomputers, as the MTC middleware requires these calls to
          start tasks.
          Some supercomputer node operating systems, for example CNK on
          Blue Gene systems, do not provide this functionality.  In
          the case of CNK, this meant that the compute node has to be rebooted
          before starting each new task.
	\item Dynamic Linking: 
	  Dynamic linking is required by many applications to load shared libraries at runtime. The absence of dynamic linking will add an undesirable barrier to compiling standard open source applications, especially those in which high performance packages written in C and FORTRAN are dynamically linked into high-productivity, front-end driver language frameworks such as Python. We see this, for example, on the Open Protein Simulator and on other chemistry codes based on its framework.
	\item POSIX-Compatible File System Access:
The OS should provide POSIX-compatible file system access to both local storage and persistent storage, since existing applications typically use POSIX libraries and assume POSIX
          semantics when accessing files.  Compute nodes---even those lacking hard disks and utilizing RAM or FLASH for local file systems---should make sufficient local file system space available to permit fast exchange of files for the inputs and outputs of MTC applications. Ideally, locally shared file systems will be made available on which collective data management strategies can implement intermediate file systems (IFSs) situated between the compute nodes and global file systems (GFSs). Such IFSs should be efficient and free of the costly locking and integrity guarantees that make GFSs problematic for efficiently managed file sharing.
          \item Intermediate Utility Nodes:
          Intermediate utility nodes provide access to intermediate I/O processors for the execution of MTC middleware, as described in experiences based on Falkon~\cite{FALKON-SC-08}. Also, utility worker nodes on the periphery of the system, with efficient interconnect access to the entire system and with sufficient resources in terms of RAM and CPU cores and speed, can run workflow managers. Such nodes should be assignable to specific user jobs and should not impact other users or jobs.

      \end{itemize}
    \item Middleware Requirements:
      \begin{itemize}
        \item Resource Provisioner:
	  A resource provisioner functions as a negotiator between the MTC middleware and the default resource manager on the supercomputer. Common resource managers include PBS, SGE, and Cobalt. A resource provisioner uses the default resource manager to both allocate and release resources. Additionally, in a dynamic approach, the resource provisioner can/may adjust the size of allocated resources according to the utilization of the allocation.
	\item Job Scheduler and Load Balancer:
	  A job scheduler dispatches jobs to available computing resources. A load balancer can be either a scheduling strategy in the job scheduler or an independent module. The load balancer avoids the starving situation when some busy compute nodes have jobs in the queue while other compute nodes are idle.
	\item Data Manager:
	  The data manager implements a data policy that decides when and where to move each type of data. The types are common input data, unique input data, intermediate data, and output data. The data manager must decide
        whether to move the input and output data synchronously at the beginning
        and end of the job or, alternatively, to prestage the input data or delay writing
        back the output data.
        The data manager also must decide the destination of a file movement: whether the local file system, intermediate file system, the global file system, or some other resource.
	\item Resilience:
	  Typically, we expect MTC middleware to provide resilience to
          intermittent faults and failures.  The middleware should be able
          to recover from the incomplete execution of a run or individual task
          failures. A run can be terminated because of hardware/software failure or concern about system utilization. Resilience requires the components of the
          system---the job scheduler, the load balancer, and the data manager---to be
           able to recover enough state to continue execution
           of the run after a failure.
	\item Programming Interface:
	  A programming interface is required so that users can flexibly generate an MTC workload. The MTC middleware programming interface could be in the form of a library that allows a user to express an MTC workflow in an existing language, or a custom programming language that expresses a workflow. Not only could a library, compiler, or interpreter execute the MTC workload, but they could also identify the various MTC patterns explored (stated in \S\ref{section:patterns}) and
 perform appropriate optimizations.
\item Flexible Scheduler Granularity:
A flexible, systemwide scheduling granularity enables portions of a larger resource allocation to be freed.
\item Communication Fabric Access Primitives:
These primitives support user access---via convenient APIs and command line interfaces---to the full power of the system's communication fabric and, in particular, to broadcast and multicast capabilities at both the file and the message level.
    \end{itemize}
\end{itemize}

\subsection{Design Choices for Middleware}
\label{sect:middleware-design}
Executing and optimizing MTC applications could, in most cases, be done
separately for each MTC application, for example by writing custom
MPI code or by using MPI library functions for collective I/O operations
or dynamic load balancing~\cite{adlb-09}.
In doing so, however, we would lose many advantages that MTC applications
gain from being represented as an abstract graph (in some cases, a DAG).  Arguably, it is best to
implement middleware that can execute a properly specified graph, for
example in the form of a Pegasus DAX file~\cite{Pegasus_2005} or a Swift script~\cite{swift-ieee09,Swift_2011},
and apply appropriate optimizations (as we will describe in \S\ref{sect:patternsupport}).

MTC middleware may provide some or all of the following functions:
\begin{itemize}
    \item Programming interface -- providing a flexible interface for users to compose MTC workloads
    \item Task scheduling and dispatch -- including load balancing
    \item I/O scheduling and coordination -- transferring data between nodes,
            and staging data in and out from a global file system
    \item Data management and caching -- tracking the availability and location of
            input and output files.  Many optimizations such as
            caching, prestaging of data, and multicasting of data could be implemented
            here.
    \item Resilience -- detecting and recovering from failures.
\end{itemize}

An MTC middleware system would typically incorporate separate components running on
different parts of a supercomputer.  Falkon, which provides task dispatch and some data
management services, has an multilevel architecture with
a central coordinator on a login node, lightweight task executors on compute nodes, and another layer of
task dispatchers that act as intermediates between the task executors and the central coordinator~\cite{falkon}.

In the remainder of this subsection, we discuss some characteristics and choices for the MTC middleware:

\begin{itemize}
  \item Programming Interface:

    The scope of the functionality provided by a MTC middleware system is
    the first and most important design decision that must be made.  
    It is also intrinsically tied to the 
    programming interface that the middleware exports.  One basic strength of MTC is that it
    can abstract away many of the implementation problems inherent in taking 
    existing codes and running them in a massively parallel way.  Hiding
    implementation details under layers of abstraction can cause problems, however, if an
    application programmer requires lower-level control. 

    For example, writing robust code to correctly and efficiently stage data between 
    file systems and nodes in a parallel computer as required for an MTC workflow
    is a significant task that can be implemented in the middleware and reused
    for many different workflows.  For most applications, this significantly
    reduces the effort required to get an application up and running in parallel
    on a supercomputer.  However, if the details of data movement are entirely 
    abstracted away by the middleware, the
    programmer will lose some control and may have difficulty implementing 
    application-specific requirements or performance improvements.

    Further examples of problems that can be abstracted away by the middleware are
    site selection, fault tolerance, job submission to different schedulers, and 
    throttling of job submissions.  Abstracting away implementation details in
    a workflow can also permit dataflow optimizations by the middleware, for example,
    grouping jobs into batches or implementing more 
    efficient communication patterns, such as multicast trees or reduction trees.

  \item Resource Provisioning:

    \begin{itemize}
      \item Dynamic vs. Static:
        With a dynamic resource provisioning strategy, the number of computational
        resources devoted to a workflow is determined by the present demands of the workflow
        and fluctuates according to the number of ready tasks available.  In contrast, with a 
        static strategy, the number of resources is fixed for the duration of the workflow.

	An obvious advantage of a dynamic strategy is that higher utilization of resources can typically be achieved. Given an scientific application with various task lengths (running time), the static resource provisioning strategy would result in a ``long tail'' at the end of the processing, when most of the resources go idle and only a small number of jobs are still running. A dynamic strategy can release resources as they become idle, thus yielding a higher utilization.  (The variable runtimes paragraph in \S\ref{sect:patternsupport} further discusses this ``long tail'' issue, which is discussed in even greater detail in~\cite{armstrong-10}.)
      \item Granularity:
	Different dynamic resource provisioning strategies will request resources in
        batches of different sizes.  

        The strategy is first constrained by the granularity provided by the parallel computer's scheduler.  Many parallel computers have hundreds to tens of thousands of nodes.  On some HPC systems, 
        allocating these nodes individually to jobs is neither efficient nor effective. 
        Thus, some vendors design their systems with a greater resource provisioning 
        granularity. For example, BG/L has a 32-node granularity, while BG/P has a 
        granularity of 64 compute nodes. TACC's Ranger has a granularity of 1 compute node.

        Given such system constraints, a resource provisioning strategy must decide how many
        nodes to request at a time.  Making many individual requests of the minimum granularity
        imposes overhead in provisioning and tracking all the requests.  It also interacts
        badly with most batch scheduling systems and scheduling policies, discouraging
        users from flooding queues with many small requests, for example by limiting the number
        of active jobs per user.  Many strategies are possible: nodes can be requested in
        increments of constant size as a workflow ramps up, in increments forming an arithmetic
        progression, or increments forming a geometric progression.

    \end{itemize}

  \item Job Scheduling and Load Balancing:

    \begin{itemize}
      \item Centralized vs. Decentralized:
	Falkon \cite{FALKON-SC-08} was originally designed as a centralized task scheduler in the grid environment, with all scheduler state stored on a single computer. When it was deployed on BG/P, the centralized architecture didn't fit the scale of the machine.  Falkon's performance dropped drastically at 1,000 compute nodes. Thus, it was altered to have a more hierarchical design, in order to better schedule large numbers of jobs on supercomputers.  Falkon's three-tier architecture consists of a submit host, a group of schedulers, and numerous workers.

      \item Push vs.~Pull:
	One principle when designing ultrascale software is avoiding any centralized point that could be a bottleneck. Falkon uses  a combined model to dispatch tasks, where the submit host pushes tasks to schedulers and workers pull tasks from the scheduler once they are free. The scheduler here is a partial centralized point in the system, as it holds a number of tasks in the queue.

	An alternative to a pull model is a push architecture. Here, the submit host and schedulers tries to push available tasks to all workers uniformly. The scheduler have a lighter load, since it no longer needs to keep the task queue in its memory.  One drawback of the push model's scheduler is that there is no guarantee that the input data for a task is ready. And there is a risk of a starving situation, where busy nodes have jobs while idle nodes don't have jobs to run. To compensate for this drawback, one could use ``work stealing'' to load balance, where a worker would ask its neighbors for tasks if it is idle.

    \end{itemize}
  \item Data Management:
    \begin{itemize}
      \item Common Input, Pull vs. Push:
	Broadcasting common input files among compute nodes could be done via a pull or a push model. The source of a common input is either the shared file system or a compute node's local storage. In a push model, the broadcast operation can be done by a direct broadcast over the network. This approach is good for broadcasting from a shared file system, since it is already supported by machine hardware because the message-passing paradigm also requires it. An alternative solution is broadcasting through a tree topology to balance the data transfer load. This approach is useful when broadcasting from one compute node to others. A pull model fetches the data on the demand of a task. An optimization of multiple fetches from the same compute nodes could be done by checking the availability of the data before fetching.

      \item Intermediate Data:
	\begin{itemize}
	\item Data-Aware Scheduling vs. Distributed Coherence Protocol:
	    Falkon has a data-aware scheduling feature. The goal is to route a task to the worker that has the input data needed for its task. This is a typical scenario in multistage scientific applications.
The advantage of this scheme is that, at some level, performance can benefit by decreasing the amount of data movement that occurs. On the other hand, it poses new challenges for the task scheduler. First, in order to route the tasks to the right workers, the scheduler has to keep a map of the global data and its location. This is feasible at some level, but eventually the scheduling will become highly inefficient as querying the map will take an unacceptable length of time. Second, if multiple input data is needed for a given task, finding the optimal will be challenging.

	    In order to avoid the disadvantages of data aware scheduling and to avoid centralized point, a distributed coherence protocol could be used. Each coherence protocol server would keep information about a certain part of the data, using a hashing function. When a worker checked for the availability of a piece of data, it could use the same hash function, then look up the data at the right server in O(1) time.
	    
	  \item Location Lookup vs. Data Store:
	    To support intermediate data caching, one could either use location lookup to find out where the data is and then move it in a peer-to-peer style or build an intermediate file system on the fly. A location lookup service would have a smaller amount of data movement since it only would need to copy from the source to the destination, whereas the data store would double the amount of data movement, with one copy needed from the source to the data store and another copy needed from the data store to the destination. By implementing a POSIX interface, the data store strategy has less coding complexity  than does the location lookup service.
	\end{itemize}

      \item Synchronized Data Movement vs. Collective Data Management:
	The data movement strategy must manage four types of data: common input data, unique input data, intermediate data, and output data. Both of the proposed strategies share common techniques for common/unique input data. They broadcast common input data to compute nodes, and workers pull unique input data from persistent storage when it is demanded by a job. Intermediate data management has been covered in the preceding paragraph. Synchronized data movement and collective data management differ when handling output data. Collective data management stores output data in a temporary shared file system, and the file system periodically backs up the data to the persistent file system. On the other hand,  synchronized data movement copies output data to the persistent file system synchronously when each job finishes.

    \end{itemize}
    
  \item Resilience:

    MTC failures can be categorized as: hardware failure, OS failure, application failure, and strategic failure. A hardware failure happens when the power is cut or hardware becomes unavailable. OS failure includes environment variable misconfiguration and out of memory. An application failure could be caused by a programming error in the application code. A strategic failure happens when the run is shut off for specific purpose, such as in ``tail chopping,'' where in order to achieve high utilization on supercomputers,a current run may be killed when 90\% of the jobs finish; then a smaller allocation of resource may be used for the rest of the jobs, restarting the killed jobs, until the whole run is completed.

    With any of the above failures, the states of the services need to be reestablished, so that the remaining jobs can resume and finish. A number of technologies exist for failure recovery, including  retry strategy and checkpointing.
Also, a group of services need to be recovered: job scheduler, data lookup service, intermediate file system. To recover the job scheduler, one could simply retry the failed and unreturned jobs. The data lookup service requires use of checkpointing, since it needs to know  both the location and the state of the intermediate data. If a different group of resources is being used after the restart,
the intermediate data on GPFS must be consistent with the information of the data lookup services, as well as information migration from larger to smaller allocation. The intermediate file system (e.g.,  MosaStore~\cite{mosastore}) should provide the functionality of checkpointing recovery, so that when  the intermediate file system is restarted on a different allocation, it can recover the data there.

\end{itemize}

\subsection{Hardware and Software Support of Patterns\label{sect:patternsupport}}
In this section, we discuss the demands each of the patterns described
in \S\ref{section:patterns} has for the MTC middleware:

\begin{itemize}

\item Control Flow Patterns:

  \begin{itemize}

  \item Parameter Sweep: Parameter sweeps are perhaps the most
    straightforward use of MTC to support. They require little more from
    software and hardware beyond the basic ability to launch parallel
    tasks on worker nodes.  However, the
    parameter sweep workload is ``flat'': all job specifications are
    available in advance.  Thus they may be efficiently organized and
    scheduled as a whole and dispatched to worker sites using high-performance messaging techniques.

  \item Iteration: Iteration requires support from the MTC
    runtime.  Some systems that assume that the DAG of tasks is static,
    such as Condor's DAGMan, cannot support iteration without augmentation.
    Another challenge posed by iteration is that 
    estimating the duration of the job may be difficult when requesting time on a
    machine, and providing an upper
    bound on the length of a job may be especially difficult: if a reservation runs
    out before the job ends, there is potential for the computational results 
    to be lost.  Fortunately, MTC supports
    the resumption of incomplete jobs, since results of each completed
    task can be written out to a file system.

  \item Task Pruning: Task pruning requires support from the MTC
    middleware to allow a running task to signal another running task
    to stop without itself terminating.

  \end{itemize}

\item Dataflow Patterns:

  \begin{itemize}

  \item Scatter: The major problem posed by the scatter pattern is that
    a very large load can be imposed on a single point in the system,
    either the GFS or the node with the data in its storage, since a
    large number of nodes will need to read the same item of data at
    the same time over the network.

    In order to mitigate this problem, a mechanism is needed in the MTC middleware for efficiently disseminating
    data to many nodes, for example, by using a multicast tree, file replication, or special hardware support for broadcast/multicast.

  \item Gather: The naive implementation of a gather would involve
    writing all the files to a GFS and then reading them back in.  A
    smarter implementation, which would require support from MTC
    middleware, would instead stage data directly from node to node,
    either through an  intermediate file system (IFS),
    created by combining the storage
     capacity of worker nodes,
    or by staging data from one task to another
    directly.  The node receiving all the data is likely to become a
    bottleneck, where the limiting factor may be network bandwidth,
    local storage availability, or compute performance. This
    bottleneck problem can be relieved by a compiler optimization.
    Instead of transferring all data to a single node where the data
    operation takes place, the compiler could generate a reduction
    tree of data operations.

    An important consideration here is whether the data is pushed to
    or pulled by the node.  If data transfers to the bottleneck node
    can be initiated by the other side (``push''), there is a risk of
    overwhelming the node.

    If data transfers can be staggered without causing delays, or if data
    can be reduced in size by processing it hierarchically, this approach is
    likely to improve performance.

  \item Pipeline: If a set of tasks is performed sequentially,
    with the output of each serving as the input to the next, this
    structure can be exploited to improve performance.  The MTC
    middleware would ideally ensure that the set of tasks was 
    assigned to the same node, allowing data to remain local to the
    node and to support efficient task dispatch so that there is
    minimal delay in executing one task once its predecessor is
    finished.  An effective way to achieve this is to dispatch the set
    of tasks together as a group.

  \item Data Reuse: If the same data is reused by different tasks
    repeatedly, the potential exists for significant, unnecessary
    strain on the GFS if the data is repeatedly fetched.

    The use of a node's local storage, or the use of an
     IFS  can significantly reduce the stress on
     the GFS when used as a cache for intermediate files or files
     from the GFS.  A data-aware task scheduler can then also
     move the computation to the data, reducing the amount of data that
     must be transferred.

  \end{itemize}

\item Other Features:

  \begin{itemize}

  \item Coordination of MPI Applications: The MTC runtime would need to
    be able to request blocks of CPUs from the MTC scheduler on which to run
    MPI tasks.  In this case, each worker in the allocated block
    would receive information from the scheduler that would allow it to
    dynamically connect to other workers, dynamically constructing the
    MPI application from its component processes.  Changes to the
    popular MPICH implementation have been made to support this mode
    of operation.

  \item Variable Runtimes: The primary problem that occurs in
    the presence of tasks with highly varying durations is that of
    efficiently utilizing available computer resources.  MTC 
    will keep worker nodes busy as long as
    sufficiently many parallel tasks are available.  Even in
    the best case, however, utilization of the available resources is
    likely to drop off significantly at the end of a job or a phase,
    as the pool of available work shrinks and only a small ``tail'' of
    running jobs remain.  This is particularly problematic when the
    distribution of task runtimes is skewed so that 
    a significant proportion of tasks have runtimes much greater
    than the mean.  If a supercomputer's allocation policy requires
    compute nodes to be requested and released in large blocks,
    this can cause unacceptably inefficient use of allocated time
    on a supercomputer.  We call this the ``trailing task''
    problem; it is discussed in detail in a
    separate paper~\cite{armstrong-10}.

    Where supported, the appropriate solution is to relinquish
    idle workers.  However, supercomputer nodes typically must be
    allocated to jobs in blocks, and it may not be possible for a job to
    relinquish nodes individually when they are no longer needed.

    Several solutions are possible.  Dispatching tasks in order of
    longest to shortest runtime is effective in achieving a better
    schedule in most cases, but it is possible only if task runtimes are
    known ahead of time.  Otherwise, reducing the number of worker
    CPUs allocated can improve utilization, but only at the cost of
    increasing time to solution.  The terms of the trade-off between
    utilization and time to solution can be improved by ``chopping off
    the tail'' of tasks remaining once some number of CPUs have fallen
    unused.  The MTC scheduler can terminate straggling tasks,
    allocate a smaller number of CPUs, and restart the work there,
    where it can complete without leaving as many resources idle.
    Specific scheduler and middleware features could be provided on a
    system such as Blue Waters to boost the effectiveness of tail chopping,
    thereby allowing scientists to obtain a quick time-to-solution
    while still using the machine efficiently. In
    particular, the ability to relinquish part of a CPU allocation
    without relinquishing the whole would allow tail chopping to
    proceed with no delay.  The ability to migrate tasks between
    worker CPUs would further facilitate this, as it would enable
    longer-running tasks to be consolidated into smaller partitions of
    a supercomputer~\cite{armstrong-10}.

  \end{itemize}

\end{itemize}

\section{Conclusions\label{sec:conclusions}}

In this report, we have discussed the concept of MTC applications, a number of
specific MTC applications, and the value of the MTC approach for
enhanced scientific productivity.  We believe that
the number of MTC applications will continue to increase and
that these applications will make up an important category of applications that will demand resources
on large-scale systems such as Blue Waters.

We believe that all the application domains we have surveyed indicate an 
important progression and trend.  The application starts out as a serial code. 
Such serial codes can almost always immediately benefit from execution as a 
many-task application because of a need to run the application on larger 
datasets and/or on an increasingly broad parameter space. Many of the algorithms 
involved also can be run in parallel with tightly coupled multiprocessing 
(e.g., MPI) or shared-memory multiprocessing (e.g., 
OpenMP). But in many cases, such fine-grained parallelism hits a ceiling of 
speedup between 1,000 and 10,000 cores, while the benefits of running 
{\em many} such application instances in parallel keeps increasing for the 
reasons above. Thus, we believe that many already parallel applications
will benefit from a hybrid model of using MTC for the 
higher-level outer loops of a program, while using fine-grained parallel processing 
in the inner loops.
We believe it will become increasingly beneficial---in terms of the scientific merits or reduced time-to-solution and hence to discovery---for Blue Waters to support such hybrid MTC-HPC applications.

In \S\ref{sec:categories} we have presented a taxonomy for identifying MTC patterns,
and in \S\ref{sec:support}, we have provided insight into core features
that the system needs to support,
focusing on what will be needed to make Blue Waters
an ideal platform for such applications.
It appears that MTC applications will be able to make basic use of the Blue Waters system without demanding any changes to its intrinsic design.
The following are some specific architectural recommendations that our study suggests for the Blue Waters system to benefit MTC applications:
\begin{itemize}
\item
Providing full Linux semantics on its compute nodes, including multiprocessing (fork/exec) and dynamic loading, so that a broad set of applications can be readily compiled and executed, and so that MTC middleware can be readily deployed.
\item Providing some (however limited)  local file system, on compute node kernels, for use in passing datasets into and out of MTC applications.
\item Providing access to intermediate I/O processors for the execution of MTC middleware, as described in experiences based on Falkon~\cite{FALKON-SC-08}.
\item Having a flexible, systemwide scheduling granularity that enables portions of a larger resource allocation to be freed.
\item Supporting user access---via convenient APIs and command line interfaces---to the full power of the system's communication fabric. 
\item Providing utility worker nodes on the periphery of the system on which workflow managers can be run with efficient interconnect access to the entire system and with sufficient resources in terms of RAM and CPU cores and speed. Such nodes should be assignable to specific user jobs and not impact other users or jobs.
\end{itemize}
We believe that all these recommendations can be met with minimal impact in the initial Blue Waters architecture, to the best of our knowledge based on current public information and experience on systems such as the BG/P, XT5, and Constellation.

Given the technical feasibility of efficient execution of MTC applications on Blue 
Waters-class systems, it thus becomes a matter of policy whether these applications 
are run on the system. We believe that the definition of ``capability systems'' should be 
interpreted by weighing the scientific merit of an application and its 
ability to efficiently use the system's resources more than the application's
specific implementation 
approach. That is, the criteria for allocation should be scientific need/merit 
and the inability to achieve that science in a timely fashion using other more readily 
available resources. The fact that an MTC application could be run on smaller 
resources while a more traditionally implemented HPC application could not is, we 
believe, a criterion that does not reflect the true value and scientific opportunity 
presented by the resource request. Another reason to better support MTC 
applications is that urgent computing situations---perhaps arising from national or global 
health, climate, weather, or defense emergencies---may require the execution of MTC 
applications at a scale far greater than that which exists under normal conditions. For 
example, using petascale resources to run BLAST might be urgent in an emergency, 
requiring a more rapid time-to-solution than could be achieve by aggregating
lower-performance resources.

In summary, MTC applications are here today, and they are increasing in number.
Blue Waters technically
can support such applications, as designed, and could better support them with some
relatively small changes.  Policies to support such applications on Blue Waters
will lead to valuable science results that would otherwise be much delayed.

\section*{Acknowledgments}

This research is part of the Blue Waters sustained-petascale computing project, which is supported by the National Science Foundation (award number OCI 0725070) and the state of Illinois. Blue Waters is a joint effort of the University of Illinois at Urbana-Champaign, its National Center for Supercomputing Applications, IBM, and the Great Lakes Consortium for Petascale Computation.
Additional support was provided by NSF under awards OCI-0944332 and OCI-1007115.

\bibliographystyle{is-unsrt}
\bibliography{MTCreport}

\begin{thebibliography}{10}
\ifx \showCODEN  \undefined \def \showCODEN #1{CODEN #1}  \fi
\ifx \showISBN   \undefined \def \showISBN  #1{ISBN #1}   \fi
\ifx \showISSN   \undefined \def \showISSN  #1{ISSN #1}   \fi
\ifx \showLCCN   \undefined \def \showLCCN  #1{LCCN #1}   \fi
\ifx \showPRICE  \undefined \def \showPRICE #1{#1}        \fi
\ifx \showURL    \undefined \def \showURL {URL }          \fi
\ifx \path       \undefined \input path.sty               \fi
\ifx \ifshowURL \undefined
     \newif \ifshowURL
     \showURLtrue
\fi

\bibitem{keyes}
David Keyes.
\newblock Exaflop/s, seriously!, 2010.
\newblock Keynote lecture for Pan-American Advanced Studies Institutes Program
  (PASI), Boston University.

\bibitem{FALKON-SC-08}
Ioan Raicu, Zhao Zhang, Mike Wilde, Ian Foster, Pete Beckman, Kamil Iskra, and
  Ben Clifford.
\newblock Toward loosely coupled programming on petascale systems.
\newblock In {\em Proc. IEEE/ACM Supercomputing 2008}, November 2008.

\bibitem{bw-Aug2011-announcement}
Timothy~Prickett Morgan.
\newblock {IBM yanks chain on `Blue Waters' super}.
\newblock {\em The Register}, 8 August 2011.
\newblock
  \url{http://www.theregister.co.uk/2011/08/08/ibm_kills_blue_waters_super/}.

\bibitem{PBS_1998}
Robert~L. Henderson and David Tweten.
\newblock Portable batch system: Requirement specification.
\newblock Technical report, NAS Systems Division, NASA Ames Research Center,
  1998.

\bibitem{falkon}
Ioan Raicu, Yong Zhao, Catalin Dumitrescu, Ian Foster, and Michael Wilde.
\newblock Falkon: a {Fast} and {Light-weight} {tasK} {executiON} framework.
\newblock In {\em Proc. IEEE/ACM Supercomputing 2007}, pages 1--12, 2007.

\bibitem{Cobalt_WWW}
Cobalt web site.
\newblock \ifshowURL {\showURL
  \path|http://trac.mcs.anl.gov/projects/cobalt|}\fi.

\bibitem{GPFS_2002}
Frank Schmuck and Roger Haskin.
\newblock {GPFS}: {A} shared-disk file system for large computing clusters.
\newblock In {\em Proc. FAST}, 2002.

\bibitem{Collective_IO_2008}
Zhao Zhang, Allan Espinosa, Kamil Iskra, Ioan Raicu, Ian Foster, and Michael
  Wilde.
\newblock Design and evaluation of a collective {I/O} model for loosely-coupled
  petascale programming.
\newblock In {\em Proc. MTAGS Workshop and SC'08}, 2008.

\bibitem{ParallelIO_2008}
Avery Ching, Kenin Coloma, Jianwei Li, Wei keng Liao, and Alok Choudhary.
\newblock High-performance techniques for parallel {I/O}.
\newblock In {\em Handbook of {P}arallel {C}omputing: {Models}, {A}lgorithms
  and {A}pplications}, chapter~35. CRC Press, 2008.

\bibitem{MPI-IO_1999}
Rajeev Thakur, William Gropp, and Ewing Lusk.
\newblock On implementing {MPI-IO} portably and with high performance.
\newblock In {\em Proc. of the Sixth Workshop on I/O in Parallel and
  Distributed Systems}, May 1999.

\bibitem{wozniak09petascalestorage}
Justin~M. Wozniak and Michael Wilde.
\newblock Case studies in storage access by loosely coupled petascale
  applications.
\newblock In {\em Proc. 4th Annual Workshop on Petascale Data Storage}, pages
  16--20, 2009.

\bibitem{DataDiffusion_2008}
I.~Raicu, Y.~Zhao, I.T. Foster, and A.~Szalay.
\newblock {Accelerating large-scale data exploration through data diffusion}.
\newblock In {\em Proc. 2008 International Workshop on Data-aware Distributed
  Computing}, 2008.

\bibitem{CMPI_2010}
J.~M. Wozniak, B.~Jacobs, R.~Latham, S.~Lang, S.~W. Son, and R.~Ross.
\newblock {C-MPI}: A {DHT} implementation for {G}rid and {HPC} environments.
\newblock Technical Report ANL/MCS-P1746-0410, Mathematics and Computer Science
  Division, Argonne National Laboratory, April 2010.

\bibitem{CloudOverview_2008}
M.A. Vouk.
\newblock Cloud computing- issues, research and implementations.
\newblock In {\em Proc. Information Technology Interfaces}, 2008.

\bibitem{CloudWorkflows_2008}
C.~Hoffa, G.~Mehta, T.~Freeman, K.~Keahey E.~Deelman, B.~Berriman, and J.~Good.
\newblock On the use of cloud computing for scientific workflows.
\newblock In {\em Proc. Scientific Workflows and Business Workflow Standards in
  e-Science}, 2008.

\bibitem{Condor_Experience_2004}
Douglas Thain, Todd Tannenbaum, and Miron Livny.
\newblock Distributed computing in practice: The {Condor} experience.
\newblock {\em Concurrency and Computation: Practice and Experience},
  17\penalty0 (2-4), 2005.

\bibitem{OSG_2007}
Ruth Pordes, Don Petravick, Bill Kramer, Doug Olson, Miron Livny, Alain Roy,
  Paul Avery, Kent Blackburn, Torre Wenaus, Frank Wurthwein, Ian Foster, Rob
  Gardner, Mike Wilde, Alan Blatecky, John McGee, and Rob Quick.
\newblock The {O}pen {S}cience {G}rid.
\newblock {\em Journal of Physics: Conference Series}, 78\penalty0 (1), 2007.

\bibitem{swift-ieee09}
Michael Wilde, Ian Foster, Kamil Iskra, Pete Beckman, Zhao Zhang, Allan
  Espinosa, Mihael Hategan, Ben Clifford, and Ioan Raicu.
\newblock Parallel scripting for applications at the petascale and beyond.
\newblock {\em Computer}, 42:\penalty0 50--60, 2009.

\bibitem{Swift_2011}
Michael Wilde, Mihael Hategan, Justin~M. Wozniak, Ben Clifford, Daniel~S. Katz,
  and Ian Foster.
\newblock Swift: {A} language for distributed parallel scripting.
\newblock {\em Parallel Computing}, pages 633--652, September 2011.

\bibitem{Pegasus_2005}
Ewa Deelman, Gurmeet Singh, Mei-Hui Su, James Blythe, Yolanda Gila, Carl
  Kesselman, Gaurang Mehta, Karan Vahi, G.~Bruce Berriman, John Good, Anastasia
  Laity, Joseph~C. Jacob, and Daniel~S. Katz.
\newblock Pegasus: A framework for mapping complex scientific workflows onto
  distributed systems.
\newblock {\em Scientific Programming}, 13, 2005.

\bibitem{Nimrod_G_2000}
David Abramson, Jon Giddy, and Lew Kotler.
\newblock High performance parametric modeling with {Nimrod/G}: Killer
  application for the global grid.
\newblock In {\em Proc. International Parallel and Distributed Processing
  Symposium}, 2000.

\bibitem{yaakoub}
Yaakoub El-Khamra and Shantenu Jha.
\newblock Developing autonomic distributed scientific applications: a case
  study from history matching using ensemble kalman-filters.
\newblock In {\em Proc. 2009 Workshop on Grids meets Autonomic Computing},
  pages 19--28, 2009.

\bibitem{CMS}
F.~W\"{u}rthwein.
\newblock Science on the grid with {CMS} at the {LHC}.
\newblock {\em Journal of Physics: Conference Series}, 125\penalty0
  (1):\penalty0 012073, 2008.

\bibitem{ligo}
B.~P.~Abbott et~al.
\newblock {LIGO}: the laser interferometer gravitational-wave observatory.
\newblock {\em Reports On Progress In Physics}, 72, 2009.

\bibitem{rosetta}
D.~Chivian, D.~E. Kim, L.~Malmstr\"{o}m, J.~Schonbrun, C.~A. Rohl, and
  D.~Baker.
\newblock Prediction of {CASP6} structures using automated {Robetta} protocols.
\newblock {\em Proteins: Structure, Function, and Bioinformatics}, 61\penalty0
  (S7):\penalty0 157--166, 2005.

\bibitem{rosetta@home}
R.~Das, B.~Qian, S.~Raman, R.~Vernon, J.~Thompson, P.~Bradley, S.~Khare, M.~D.
  Tyka, D.~Bhat, D.~Chivian, D.~E. Kim, W.~H. Sheffler, L.~Malmstr\"{o}m, A.~M.
  Wollacott, C.~Wang, I.~Andre, and D.~Baker.
\newblock Structure prediction for {CASP7} targets using extensive all-atom
  refinement with {Rosetta@home}.
\newblock {\em Proteins: Structure, Function, and Bioinformatics}, 69\penalty0
  (S8):\penalty0 118--128, 2007.

\bibitem{robetta}
David~E. Kim, Dylan Chivian, and David Baker.
\newblock {Protein structure prediction and analysis using the Robetta server}.
\newblock {\em Nucleic Acids Research}, 32\penalty0 (suppl 2):\penalty0
  W526--W531, 2004.

\bibitem{Greeley2006}
J.~Greeley, J.~Jaramillo, J.~Bonde, I.~Chorkendorff, and J.~Norskov.
\newblock Computational high-throughput screening of electrocatalytic materials
  for hydrogen evolution.
\newblock {\em Nature Materials}, 5:\penalty0 909--913, 2006.

\bibitem{Greeley2009}
J.~Greeley and J.~Norskov.
\newblock Combinatorial density functional theory-based screening of surface
  alloys for the oxygen reduction reaction.
\newblock {\em Journal of Physical Chemistry C}, 113:\penalty0 4932--4939,
  2009.

\bibitem{Munson2010}
Todd Munson, February 18, 2010.
\newblock personal communication.

\bibitem{raicu06astroportal}
Ioan Raicu, Ian Foster, Alex Szalay, and Gabriela Turcu.
\newblock {AstroPortal}: A science gateway for large-scale astronomy data
  analysis.
\newblock In {\em Proc. {TeraGrid} Conference 2006}, 2006.

\bibitem{ptmap09}
Yue Chen, Wei Chen, Melanie~H. Cobb, and Yingming Zhao.
\newblock {PTMap: A sequence alignment software for unrestricted, accurate, and
  full-spectrum identification of post-translational modification sites}.
\newblock {\em Proc. National Academy of Sciences}, 106\penalty0 (3):\penalty0
  761--766, 2009.

\bibitem{hocky-oops-09}
G.~Hocky, M.~Wilde, J.~Debartolo, M.~Hategan, I.~Foster, T.R. Sosnick, and K.F.
  Freed.
\newblock Towards petascale ab initio protein folding through parallel
  scripting, April 2009.
\newblock Argonne National Laboratory, Mathematics and Computer Science
  Division preprint ANL/MCS-P1612-0409.

\bibitem{IBM-DOCK-08}
Amanda Peters, Marcus~E. Lundberg, P.~Therese Lang, and Carlos~P. Sosa.
\newblock High throughput computing validation for drug discovery using the
  {DOCK} program on a massively parallel system.
\newblock In {\em Proc. 1st Annual Midwest Symposium on Computational Biology
  and Bioinformatics}, September 2007.

\bibitem{montage1}
Joseph~C. Jacob, Daniel~S. Katz, G.~Bruce Berriman, John~C. Good, Anastasia~C.
  Laity, Ewa Deelman, Carl Kesselman, Gurmeet Singh, Mei-Hui Su, Thomas~A.
  Prince, and Roy Williams.
\newblock Montage: a grid portal and software toolkit for science-grade
  astronomical image mosaicking.
\newblock {\em International Journal of Computational Science and Engineering},
  4\penalty0 (2):\penalty0 73--87, 2009.

\bibitem{montage2}
Daniel~S. Katz, Joseph~C. Jacob, G.~Bruce Berriman, John Good, Anastasia~C.
  Laity, Ewa Deelman, Carl Kesselman, and Gurmeet Singh.
\newblock A comparison of two methods for building astronomical image mosaics
  on a grid.
\newblock In {\em Proc. 2005 International Conference on Parallel Processing
  Workshops}, pages 85--94, 2005.
\newblock \showISBN{0-7695-2381-1}.

\bibitem{rendell-sociallearning-09}
L.~Rendell, L.~Fogarty, and K.~N. Laland.
\newblock Rogers' paradox recast and resolved: Population structure and the
  evolution of social learning strategies.
\newblock {\em Evolution}, 64\penalty0 (2):\penalty0 534--548, 2009.

\bibitem{rendell-sociallearning-10}
L.~Rendell, R.~Boyd, D.~Cownden, M.~Enquist, K.~Eriksson, M.~W. Feldman,
  L.~Fogarty, S.~Ghirlanda, T.~Lillicrap, and K.~N. Laland.
\newblock Why copy others? {Insights} from the social learning strategies
  tournament.
\newblock {\em Science}, 328\penalty0 (5975):\penalty0 208--213, April 9, 2010.

\bibitem{rendell-email-10}
Luke Rendell, 2010.
\newblock personal communication.

\bibitem{blast-1990}
Stephen~F. Altschul, Warren Gish, Webb Miller, Eugene~W. Myers, and David~J.
  Lipman.
\newblock Basic local alignment search tool.
\newblock {\em Journal of Molecular Biology}, 215:\penalty0 403--410, 1990.

\bibitem{mpiBLAST-SC-08}
H.~Lin, P.~Balaji, R.~Poole, C.~Sosa, X.~Ma, and W.~Feng.
\newblock Massively parallel genomic sequence search on the {Blue} {Gene}/{P}
  architecture.
\newblock In {\em Proc. IEEE/ACM Supercomputing 2008}, November 2008.

\bibitem{CIM-EARTH-10}
J.~Elliott, I.~Foster, K.~Judd, E.~Moyer, and T.~Munson.
\newblock {CIM-EARTH:} {Philosophy}, {Models}, and {Case} {Studies}, Feb 2010.
\newblock Argonne National Laboratory, Mathematics and Computer Science
  Division preprint ANL/MCS-P1710-1209.

\bibitem{nugent-synapps-08}
P.~Nugent, R.~Thomas, and G.~Aldering.
\newblock Optimizing type {Ia} supernova follow-up in future dark energy
  surveys.
\newblock {\em Journal of Physics: Conference Series}, 125\penalty0
  (1):\penalty0 012011, 2008.

\bibitem{branch-synow-07}
David Branch, Jerod Parrent, M.~A. Troxel, D.~Casebeer, David~J. Jeffery,
  E.~Baron, Wesley Ketchum, and Nicholas Hall.
\newblock Probing the nature of type {I} supernovae with {SYNOW}.
\newblock {\em AIP Conference Proceedings}, 924\penalty0 (1):\penalty0
  342--349, 2007.

\bibitem{deem-2009}
Michael~W. Deem, Ramdas Pophale, Phillip~A. Cheeseman, and David~J. Earl.
\newblock Computational discovery of new zeolite-like materials.
\newblock {\em The Journal of Physical Chemistry C}, 113:\penalty0
  21353--21360, 2009.

\bibitem{walker-07}
E.~Walker, David~J. Earl, and Michael~W. Deem.
\newblock How to run a million jobs in six months on the {NSF} {TeraGrid}.
\newblock In {\em Proc. TeraGrid Conference 2007}, 2007.

\bibitem{fMRI-1994}
A.~R. McIntosh, C.~L. Grady, L.~G. Ungerleider, J.~V. Haxby, S.~I. Rapoport,
  and B.~Horwitz.
\newblock Network analysis of cortical visual pathways mapped with {PET}.
\newblock {\em Journal of Neuroscience}, 14:\penalty0 655--666, 1994.

\bibitem{fMRI-1995}
F.~Gonzalez-Lima and A.~R. McIntosh.
\newblock Analysis of neural network interactions related to associative
  learning using structural equation modeling.
\newblock {\em Mathematics and Computers in Simulation}, 40:\penalty0 115--140,
  1995.

\bibitem{henry-09}
Christopher~S. Henry, Fangfang Xia, and Rick Stevens.
\newblock Application of high-performance computing to the reconstruction,
  analysis, and optimization of genome-scale metabolic models.
\newblock {\em Journal of Physics: Conference Series}, 180\penalty0
  (1):\penalty0 012025, 2009.

\bibitem{henry-10}
Christopher~S. Henry, Matthew DeJongh, Aaron~A. Best, Paul~M. Frybarger, Ben
  Linsay, and Rick~L. Stevens.
\newblock High-throughput generation, optimization and analysis of genome-scale
  metabolic models.
\newblock {\em Nature Biotechology}, 28\penalty0 (9):\penalty0 977--982,
  September 2010.
\newblock \showISSN{1087-0156}.

\bibitem{seed-manifesto-04}
Ross Overbeek.
\newblock The project to annotate the first 1000 sequenced genomes, develop
  detailed metabolic reconstructions, and construct the corresponding
  stoichiometric matrices.
\newblock \url{http://www.nmpdr.org/FIG/html/1KG_update.html}.

\bibitem{henry-email-10}
Christopher~S. Henry, 2010.
\newblock personal communication.

\bibitem{adlb-09}
E.~L. Lusk, S.~C. Pieper, and R.~M. Butler.
\newblock More scalability, less pain: A simple programming model and its
  implementation for extreme computing.
\newblock {\em SciDAC Review}, 17:\penalty0 30--37, spring 2010.

\bibitem{armstrong-10}
Timothy~G. Armstrong, Zhao Zhang, Daniel~S. Katz, Michael Wilde, and Ian
  Foster.
\newblock Scheduling many-task workloads on supercomputers: Dealing with
  trailing tasks.
\newblock In {\em Proceedings of 3rd Workshop on Many-Task Computing on Grids
  and Supercomputers}, 2010.

\bibitem{mosastore}
Samer Al-Kiswany, Abdullah Gharaibeh, and Matei Ripeanu.
\newblock The case for a versatile storage system.
\newblock {\em SIGOPS Oper. Syst. Rev.}, 44\penalty0 (1):\penalty0 10--14,
  2010.
\newblock \showISSN{0163-5980}.

\end{thebibliography}

The submitted report has been partially created by UChicago Argonne, LLC, Operator of Argonne National Laboratory (``Argonne''). Argonne, a U.S. Department of Energy Office of Science laboratory, is operated under Contract No. DE-AC02-06CH11357. The U.S. Government retains for itself, and others acting on its behalf, a paid-up nonexclusive, irrevocable worldwide license in said article to reproduce, prepare derivative works, distribute copies to the public, and perform publicly and display publicly, by or on behalf of the Government.

\end{document}